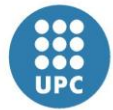 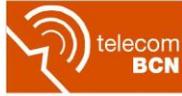

**ESCOLA TÈCNICA SUPERIOR d'Enginyeria de Telecomunicació de Barcelona**

UNIVERSITAT POLITÈCNICA DE CATALUNYA

# PROJECTE FINAL DE CARRERA

## Improvement of algorithms to identify transportation modes for MobilitApp, an Android Application to anonymously track citizens in Barcelona

*Degree:* Telecommunications Engineering

*Author:* Gerard Marrugat Torregrosa

*Director:* Mónica Aguilar Igartua

Co-Director: Silvia Puglisi

*Date: May 2016*

*Dedicated to all friends, colleagues and family who are always supporting me*

# Resum del projecte

MobilitApp (_http://mobilitapp.noip.me/_) es una aplicació Android que té com a objectiu obtenir dades de mobilitat dels ciutadans de l'Àrea Metropolitana de Barcelona. En aquest projecte ens hem centrat en la recerca d'algoritmes més fidels i robustos en la decisió del mitjà de transport utilitzat per l'usuari basant-nos en tècniques d'accelerometria.

L'algoritme desenvolupat agafa dades de l'acceleròmetre i giroscopi del mòbil i les guarda en un fitxer que posteriorment es enviat al servidor. Aquest procediment es realitzà en segon pla sense interferir en l'activitat principal de l'aplicació. Les dades recollides i emmagatzemades al servidor han sigut processades i utilitzades per a realitzar un anàlisi del comportament del patró de mobilitat que té cada medi de transport. El resultat obtingut han sigut uns paràmetres que permetran modelar l'activitat de cada mitjà i dissenyar un algoritme de detecció de la modalitat de transport que utilitzi informació obtinguda a partir dels sensors del mateix mòbil.

L'aplicació segueix realitzant la seva funcionalitat principal, la monitorització de les dades de mobilitat dels ciutadans. En futures versions s'integrarà en local la solució a la detecció del mitjà de transport proposada, d'aquesta manera l'aplicació farà ús d'informació que pot obtenir del mateix dispositiu i així no dependre d'altres serveis.

# Resumen del proyecto

MobilitApp (*http://mobilitapp.noip.me/*) es una aplicación Android que tiene como objetivo obtener datos de movilidad de los ciudadanos del Área Metropolitana de Barcelona. En este proyecto nos hemos centrado en la búsqueda de algoritmos más fiables y robustos en la decisión del medio de transporte utilizado por el usuario basándonos en técnicas de acelerometría.

El algoritmo desarrollado coge datos del acelerómetro y el giroscopio del móvil y los guarda en un fichero que posteriormente es enviado al servidor. Este procedimiento se realiza en segundo plano sin interferir en la actividad principal de la aplicación. Los datos recogidos y almacenados en el servidor han sido procesados y utilizados para realizar un análisis del comportamiento del patrón de movilidad que presenta cada medio de transporte. El resultado obtenido han sido unos parámetros que permitirán modelar la actividad de cada medio y diseñar un algoritmo de detección de la modalidad de transporte que utilice información obtenida a partir de los sensores del propio móvil.

La aplicación sigue realizando su funcionalidad principal, la monitorización de los datos de movilidad de los ciudadanos. En futuras versiones se integrará en local la solución a la detección del medio de transporte propuesta, de esta manera la aplicación hará uso de información que puede obtener del mismo dispositivo y así no depender de otros servicios.

# Abstract

MobilitApp (*http://mobilitapp.noip.me/*) is an Android application whose objective is to obtain mobility data from the citizens of the metropolitan area of Barcelona. The current project is based on the research of more trustful and stronger transport decision algorithms using advantages of accelerometry techniques.

The developed algorithm reads data from the mobile's accelerometer and gyroscope and writes it in a file that is afterwards sent to the server. This process is executed in background without interfering in the main application activity. Collected data has been processed and used to analyze the behaviour of the mobility pattern of the distinct transport modalities. The obtained result has been parameters which allow us to configure a model for each mean activity and designing a transport mode detection algorithm which would use information obtained from the mobile´s own sensors.

MobilitApp is still executing its main functionality, monitoring mobility data from the citizens. In future versions the solution proposed to detect the transport systems will be integrated into the application, in this way the app will work with information obtained from the device and will not depend on any other services.

# Index

# List of figures

# List of tables

# Glossary of acronyms

**AMB** Àrea Metropolitana de Barcelona

**API** Application Programming Interface

**APK** Android Application Package

**APP** Application

**ASL** Accelerometer Sensor Listener

**ATM** Autoritat del Transport Metropolità

**C0₂** Carbon Dioxide

**CSV** Comma-Separated Values

**DGT** Dirección General de Tráfico

**DNS** Domain Name Server

**GPS** Global Positioning System

**GUI** Graphical User Interface

**HTTP** HyperText Transfer Protocol

**ICT** Information and Communications Technology

**JSON** JavaScript Object Notation

**MySQL** My Structured Query Language

**PHP** Pre Hypertext Processor

**RMSE** Root Mean Square Error

**SDK** Software Development Kit

**SSH** Secure Shell

**STD** Standard Deviation

**UPC** Universitat Politècnica de Catalunya

**VNC** Virtual Network Computing

**Wi-Fi** Wireless-Fidelity

**XML** eXtensible Markup Language

**WPS** Wi-Fi based Positioning System

# 1.  Introduction

In this project, we are going to develop an Android application called MobilitApp [1] to obtain data regarding the mobility of citizens in Barcelona. In collaboration with Autoritat del Transport Metropolità, this project´s main objective is to use this data to determine mobility patterns that could be used to improve current transportation infrastructure.

Android Platform is used to develop MobilitApp for two main reasons:

- **Large Audience:** According to [2] in Spain at December 2014, 83% of the devices use Android.
- **Google APIs:** Google provides developers large amount of APIs (Application Programming Interface) to add different features to Apps. MobilitApp uses: Maps, Places, Directions, Location and Activity Recognition.

This study is part of the larger EMRISCO [3] project and will help ATM of Barcelona providing mobility data to improve the current transportation infrastructure.

In previous projects [4][5] the transport mode detection is still failing due to the use of inaccurate Google APIs, it is not able to distinguish between the different modalities. In this current project we have focused our efforts on the development of a solution which will correct those inaccuracies and furthermore will not depend on external services for the activity detection task.

The main objectives of this Master Thesis work are (1) to analyse the mobility patterns for each mean of transport based on accelerometer's data values gathered from the citizen's smartphones [6], and (2) to propose a classification model for the detection task. Future releases of MobilitApp would include this improvement.

In order to achieve these goals we have developed an application that listens the motion sensors in the mobile phone (i.e., accelerometer and gyroscope) and sends the collected data to a server. This tool has been added to MobilitApp as a new feature which runs in background and helps us to obtain information from the different users' activities. Once the collecting process is finished, an analysis of the stored information is done. The results of this analysis would be a group of parameters which allow us to distinguish between the different transportation modalities, bus, metro, train, car and motorbike.

This document is structured with six chapters plus references and three annexes. The Chapter 2 introduces the context where is located our research.
The next ones, Chapters 3 and 4, describes the design and implementation process of our tool, giving details on how it runs and how it interacts with MobilitApp.
Later, on Chapter 5 we report the results of the accelerometer data analysis and explain the methodology used to achieve it as well.

After that, Annex A explains how the server has been set up step by step. Annex B shows the extra features added to the project like the web page and the promotional video, and finally Annex C lists many issues found during the development.

MobilitApp final degree project (App and Report) can be found in: [7]

# 2. Mobility Area

## 2.1 Smart Cities and Mobility

### 2.1.1 Smart City Concept

The Smart city concept is an urban and technological development focused on satisfying needs and improving life quality among the inhabitants of urban areas by using highly efficient and technical services. Some of the main areas included in the smart city concept are public services, socio-cultural environment, medical and health considerations and sustainability. The idea of being a smart city is not only a goal to achieve, but it is also a long-term project which is constantly changing and arranging to the current society requests.

Technology has an important role in a smart city, but it is not the only mechanism used to build solutions in a smart city. Digital technologies and Information and Communication Technologies (ICT) give support to those offered services by creating tools like: mobile applications, Open Data storage, real time public transport monitor, public WiFi networks…

### 2.1.2 Smart Mobility

#### *Solutions on moving around the city*

Moving around the city presents many problems, regardless whether it is with public or private transport. For this reason public organizations are developing solutions on mobility fields. Many problems are directly related with citizens' daily life (socio-cultural environment); others affect the environment (sustainability). The list below shows some of them:

- Traffic Jams
- Variations in the public transport timetable
- Constructions on the road
- Traffic accidents

**Citizens' problems**

- High environmental pollution level
- Noise pollution

**Environment's problems**

Smart Mobility is one of the tools to achieve a sustainable city development. An efficient and comfortable transport system and reducing the $CO_2$ emissions are the main objectives.

Services and initiatives which improve the transport system are:

- **Smart Parking:** monitoring the parking areas and providing information about the number of unoccupied places and their location.
- **Car Sharing:** a fleet of cars is provided for sharing in certain location. The idea consists on reserving the car during the required hours, picking it up from the stipulated site and returning it at the same place. Using a card provided by the entities that offer this kind of services, users are able to open the vehicles. Car sharing proposal also reduces the $CO_2$ emissions.

- **Taxi Reservation Service:** a taxi reservation at a certain time in a certain place is done with a mobile application. The service informs the user about the price and allows him to pay via app.

Carbon gas emissions that pollute the environment are caused above all by the fossil fuel consumption in vehicles. Innovations and investigations that avoid increasing pollution levels:

- **E-Mobility:** electric motor engine vehicles are the alternative to conventional fuel consumption. This innovative and green solution is supported by the public administrations which are renewing the public services vehicles (garbage trucks, public transport…), placing charging stations around the city and subsiding the purchase of private electric vehicles.
- **Biofuels:** biomass can be converted into liquid fuels and be used in transportation proposals. Alcohols obtained by the fermentation of carbohydrates, vegetable oil, animal fat or recycled cooking grease are examples of biofuels. It is commonly mixed with gasoline or other fossil fuels to reduce vehicle emissions, but it is also used in its pure form as a renewable energy.
- **Hydrogen:** hydrogen in its pure form, $H_2$, reacts with oxygen, $O_2$, to form water and release energy. Therefore hydrogen fuel is a zero-emission fuel not considering water as an emission. Many companies are working hard to develop cells and batteries that can efficiently take advantage of hydrogen potential.

## 2.2    Mobility Applications. State of the Art

MobilitApp, the mobile application that we are developing, and the other applications described in the lines below are placed in the socio-cultural environment area of a smart city, specifically in the mobility field. In this section we want to give an overview of which services can find the citizens in the mobility apps field, and compare it to our work.

### 2.2.1 MobilitApp

*Tracking the citizens´ daily journeys*

#### 2.2.1.1    Description and Functionality

MobilitApp is a mobile application whose main purpose is to collect mobility data from the citizens of Barcelona and eventually analyze it to determine mobility patterns that could be used to improve ongoing transportation infrastructures. A web application is implemented to visualize the routes and commutes of the users. The information can be filtered by age, activity, gender, time and date.

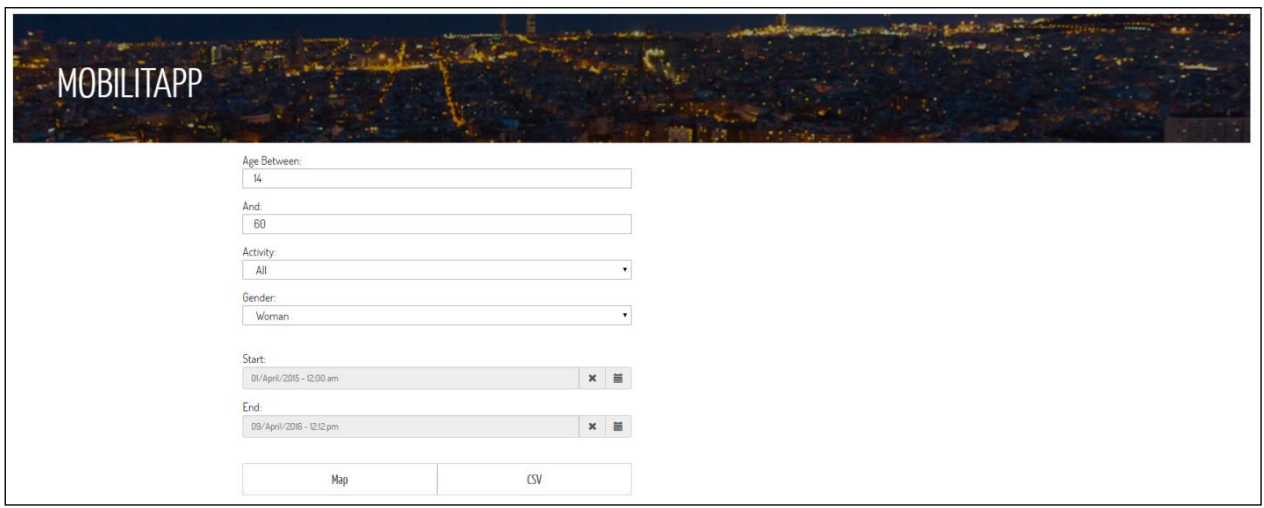

**Fig. 2-1: Web Application**

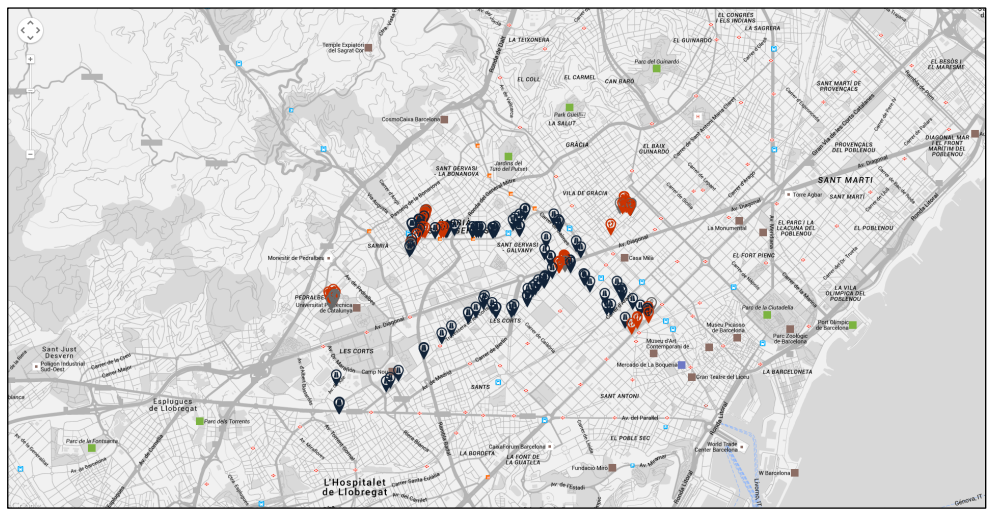

**Fig. 2-2: Consulting DataBase**

The app also gives real-time information about the traffic state and incidents on the road, allowing the citizens to take decisions on their journey.

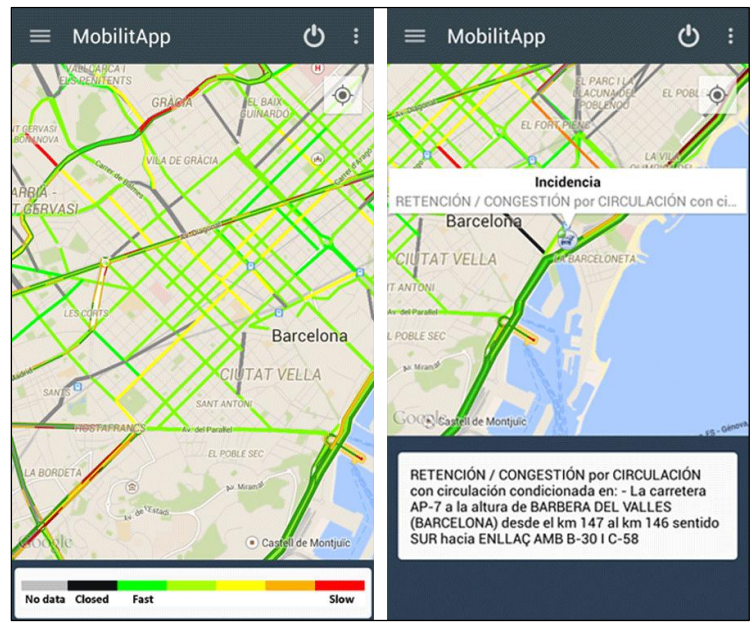

**Fig. 2-3: Real time traffic state information**

### 2.2.2 Moovit

*Riding the city smart*

#### 2.2.2.1 Description and Functionality

Moovit is a personal transit assistant, it guides users to their destinations, lets them know the closest stations and the changes in arrival and departure times.

The provisions of the service are focused on helping users to take decisions on their itineraries. They introduce the destination on the application´s search bar, *Home* screen, and a list of options with timing, transport lines and stations information is returned. Furthermore, when users are on route, the feature *En Route* allows them to visualize the estimated arrival time updated even if the final location has changed. There is a section called *Favorites* where home, work or places of interest are saved, and in case users would like to imply it, to habitually notify them the best route options.

In addition to the features explained above, Moovit attends to create a big community where users can share information about delays on schedule, vehicles capacity, events in the city and give their opinion about the services.

#### 2.2.2.2 Similarities

- Although the way is not the same, Moovit and MobilitApp report information about incidences.

- Moovit advices users about the best routes for their ways; MobilitApp just obtains data from these routes.
- Moovit is focused on public transport; MobilitApp considers private vehicles as well.
- Moovit does not inform about the traffic state; MobilitApp does.
- Moovit users are able to notify incidences on their routes.

## 2.2.3  Citymapper

*Public transport, walking and cycling directions*

### 2.2.3.1    Description and Functionality

Citymapper has been built for commuters and their daily needs. The main screen shows three tabs; *City*, *Go* and *Nearby*. The app starts on the *Go* tab, where the users can set their daily destinations; home, office, university…, to receive information regularly about the best ways to follow or simply search for an unusual route. With this app users can share the location of a future meeting spot via instant message services; this is a new feature in this kind of applications. In the *City* tab a list of all city lines is displayed and messages from public transport agencies on social networks are copied to inform about incidents or special services.

Finally, *Nearby* shows to the users the closer stations of the different kind of transport and its next departure time.

Citymapper is a web application too. In addition to the information supplies the users are able to print their itineraries before starting their ways.

### 2.2.3.2    Similarities

- Citymapper and MobilitApp provide data about incidences but in different approaches.
- Citymapper also uses a web site to provide its service; MobilitApp collects data that can be visualized in a web page.

### 2.2.3.3    Differences

- Citymapper suggests the best itineraries; MobilitApp monitors the routes.
- Citymapper shares info for public transport users, including taxi into the public transport collective; MobilitApp also takes into account private vehicles.
- Citymapper does not inform about the traffic state; MobilitApp does.
- Citymapper integrates social networks into it; MobilitApp does not.

### 2.2.4 Waze

*Outsmarting traffic drivers community*

#### 2.2.4.1  Description and Functionality

Waze mobile application works like a big community of drivers where they share information about the state of the traffic and unexpected events on the road. It is focused just on private vehicles; public transport reports are not available in the app services. Locations of interest can be registered in the app to receive notifications in real-time that affect users' commutes. All the information shared is not exclusively about the traffic; users can inform about police control points or petrol station prices.

Waze service is also offered in a web page, but just visualizing the information reported by the other users.

#### 2.2.4.2  Similarities

- In addition to the mobile application, Waze uses a web site to provide its service; MobilitApp too.

#### 2.2.4.3  Differences

- Waze only considers private vehicles, MobilitApp reports public transport information as well.
- Waze suggests the best itineraries; MobilitApp monitors the routes.
- Waze users are able to give information in real-time.

## 2.3    Conclusion about Smart Solutions on Mobility Applications

The main objective of this application group is to provide the fastest path of a trip, and to build a community of users helping each other. MobilitApp is a tool to study new alternatives that will improve people´s commutes, but it is not providing any direct benefits to the person who has the app installed. It is clear that sharing information between the users of the same place is a good way to obtain real-time data of what is happening in the city. Social networks could play an important role in this task in a near future. In addition to the mobile application these services also have a web site. Having additional platforms is a good option to show all the information related to the app and another way to provide data to people, which would mean an increasing number of users running the application services.

# 3. Accelerometer Sensor Listener

## 3.1 MobilitApp Architecture

MobilitApp is developed over Android platform which allows developers using Google APIs to add different features to applications. MobilitApp uses these Application Programming Interfaces (APIs): Maps [8], Places [9], Directions [10], Location Services [11] and Activity Recognition [12]. In this Chapter we are going to describe these Google APIs and explain the two most important tasks of MobilitApp application: Location and Activity Recognition.

### 3.1.1 APIs. Application Programming Interfaces

#### 3.1.1.1 Description and benefits of its use

Google has developed a set of features which allow communication with Google Services, these features are the Applicacion Programming Interfaces (APIs). Developers can use these APIs, take benefit of this functionality and not spending time in the development of the service´s algorithm.

#### 3.1.1.2 Activity Recognition API. Weaknesses of the feature

Activity Recognition API provides a low power consumption mechanism that can provide periodic updates of detected types by using low power sensor, basically accelerometer. Despite these advantages many problems related with Activity Recognition current service have been found using MobilitApp.

- Activity Recognition API does not distinguish between the different motorized transports, even it does detect activity as *unknown* instead of *vehicle*.
- It requires network connection and if we are in an isolated area this service cannot be offered.
- It is an external service, the data provided is not information that comes from the mobile device.
- Using API Services involves sharing users' information with Google which means a loss in privacy.

Later these issues will be resumed to expose a proposal which could solve them.

#### 3.1.1.3 Places and Directions API. How MobilitApp distinguishes among the different modalities

As it has been mentioned before, Activity Recognition API does not differentiate between motorized transports, and then the method used to recognize the type of public transport used is the following (the current mechanism is focused on public transport).

For this purpose two more Google APIs are used: Directions API and Places API. In summary Places API returns information about places (e.g. establishments, metro station…) and Directions API searches for directions of several transportation modes. Furthermore, two databases of train and metro stations are used to complement the algorithm. The databases contain the line name, station name (in the case of the metro), latitude and longitude, etc.

```
"metro":[
        …
{
"id":"422",
"line":"L5",
"name":"Hospital Clinic",
"connections":"",
"lat":"41.3886531870785",
"lon":"2.15112142588565"
},
{
"id":"370",
"line":"L3",
"name":"Fontana",
"connections":"",
"lat":"41.4020175545979",
"lon":"2.15305426943399"
},
        …
```

```
"renfe":[
        …
{
"line":"R4",
"lat":"41.3145934553543",
"lon":"1.66019560008191"
},
{
"line":"R4",
"lat":"41.3447755935616",
"lon":"1.70355768548077"
},
{
"line":"R4",
"lat":"41.3705005020672",
"lon":"1.72496778423192"
},
        …
```

**Metro**

Places API is used by the Metro detection activity. In the underground there are neither Wi-Fi spots nor sky visibility, then WPS and GPS mechanism are useless. The algorithm works as follows: when a citizen enters into the metro Places API will find the nearest station to his current location, once the metro journey is finished the closest station will be found by Places API again. The Google API Service returns the station names, then the next step is searching these names in the metro database mentioned before. When both stations are identified the metro journey segment is built.

**Bus, Tram and Train**

Before uploading the file which contains the route done by the user some validations are done. If a vehicle activity has been detected, the recognition algorithm checks if the route coincides with bus or tram routes considering beginning and end points and departure and arrival times. This can be done with the use of Directions API.

Train Recognition is done if the Bus/Tram recognition returns nothing. The train method does not use APIs, it simply searches in the train database the nearest train stations for origin and destination points and if the distance is smaller than 50 meters and both stations belong the same line activity is changed to train. Places API does not return information about trains stations and therefore train recognition is done in this way.

This was the first proposal to distinguish among the transport modalities and the result was not successful. The main detected issues:

- The algorithm designed seldom works efficiently, journeys of the most common public transportation modes inside the city, metro and bus, are not detected.

- It is not an accurate method because is based in features which are not related with the activity done.
- It is an external service, the algorithm works with data that does not come from the mobile device.
- Using API Services involves sharing users' information with Google which means a loss in privacy.

## 3.1.2 MobilitApp Workflow

### 3.1.2.1 Location Service

In Location Service section the procedure of how application obtains the current location of users is described step by step, detailing which classes take part in it. The explanation covers from initialization to update location.

#### LOCATION SERVICE EXECUTION PROCESS

Location Service is initiated every time MobilitApp is opened. Inside MainFragment, main activity's fragment application, the Requester class is instantiated, from where a call to Location Service will be generated.

**MAIN FRAGMENT CLASS**

Create an instance to Requester class inside *onActivityCreated(…)* method.

```
public void onActivityCreated(Bundle savedInstanceState) {
...
    mrequesterLOC = new Requester(getActivity());
...
}
```

Before requesting location updates Google Play Services and WiFi connection are checked.

```
public void onResume() {
...
    checkGooglePlayServices();
    checkWIFI();
...
    isProviderEnabled();
}
```

The next *requestUpdates(…)* statement does a request for location updates without using GPS. The first argument indicates the type of service, if "LOC" is written Location updates are initialized and Activity updates if it is "AR". The second argument denotes if GPS is needed or not.

```
private void isProviderEnabled() {
...
    mrequesterLOC.requestUpdates("LOC", false);
...
}
```

To use an API service a Google API Client [13] is required. In Requester class all the methods related with the use of APIs are implanted.

When *requestUpdates(…)* is called, the connection process to Google Services will be immediately initialized.

```java
public void requestUpdates(String type, boolean gps) {

    mType = type;
    mgps = gps;

    requestConnection();
}

private void requestConnection() {
    getClient().connect();
}
```

*getclient()* is the method which will build a Google API client if a connection to API´s services is required.

```java
private GoogleApiClient getClient() {
    if (myClient == null) {

            myClient = new GoogleApiClient.Builder(mContext).addApi(ActivityRecognition.API)
            .addApi(LocationServices.API).addConnectionCallbacks(this)
            .addOnConnectionFailedListener(this)
            .build();
    }
    return myClient;
}
```

Once the client is created *requestConnection()* will call *connect()* statement, then a call-back is generated in *onConnected(…)* method.

```java
public void onConnected(Bundle connectionHint) {

    continueRequestActivityUpdates(mType);
}
```

*continueRequestActivityUpdates(…)* is the recall mentioned before, here all the updates are set and finally the client is disconnected.

```
private void continueRequestActivityUpdates(String type) {
    if (type.equalsIgnoreCase("AR")) {
        ActivityRecognition.ActivityRecognitionApi.requestActivityUpdates(getClient(),
            AppUtils.UPDATE_INTERVAL_ACTIVITY, createRequestPendingIntent(type));
    } else {
        LocationServices.FusedLocationApi.requestLocationUpdates(getClient(), createLR(mgps),
            createRequestPendingIntent(type));
    }
    // Disconnect the client
    requestDisconnection();
}
```

*PendingIntent* [14] is the most efficient way to request updates in background, this is created in *createRequestPendingIntent (String type)* method. The only parameter needed, *type*, is to differentiate between the two APIs.

```
private PendingIntent createRequestPendingIntent(String type) {

    PendingIntent pendingIntent;
    // If the PendingIntent already exists
    if (null != getRequestPendingIntent()) {

        // Return the existing intent
        return mActivityRecognitionPendingIntent;

    // If no PendingIntent exists
    } else {

        if (type.equalsIgnoreCase("AR")) {

            Intent intent = new Intent(mContext, ActivityRecognitionService.class);

            pendingIntent = PendingIntent.getService(mContext, 1, intent, PendingIntent.
                FLAG_UPDATE_CURRENT);

            setRequestPendingIntent(pendingIntent);
        } else {

            Intent intent = new Intent(mContext, LocationService.class);

            pendingIntent = PendingIntent.getService(mContext, 0, intent, PendingIntent.
                FLAG_UPDATE_CURRENT);

            setRequestPendingIntent(pendingIntent);
        }
        return pendingIntent;
    }

}
```

Once the type is tested an intent to one of our two services is created; LocationService and ActivityRecognitionService classes, and then this intent is passed to the pending intent generated from the static method *getService(…).*

With these lines of code Location Updates have been achieve successfully. These pending intents will call LocationService class periodically to set the new location, but how frequently are these ones generated? *createLR(…)* method establishes the interval. Depending on the circumstances the interval could change. In normal conditions the location is provided by WPS and updates are done every 20 seconds, but if the algorithm obtains a group of bad locations consecutively the GPS will be enabled and the interval will be set to 8 seconds to recover the current location as soon as possible.

```
private LocationRequest createLR(boolean gps) {
    LocationRequest locationRequest = null;
    if (!gps) {
        locationRequest = LocationRequest.create().setPriority(LocationRequest.
        PRIORITY_BALANCED_POWER_ACCURACY)
        .setInterval(AppUtils.UPDATE_INTERVAL_LOCATION)
        .setFastestInterval(AppUtils.FASTEST_INTERVAL_LOCATION);
    } else {
        if (isGPSEnabled)
            locationRequest = LocationRequest.create().setPriority(LocationRequest.
            PRIORITY_HIGH_ACCURACY)
            .setInterval(AppUtils.UPDATE_INTERVAL_GPS);
    }
    return locationRequest;
}
```

**LOCATION SERVICE CLASS**

Every 8 or 20 seconds (depending if GPS is needed) the *onStartCommand()* method of Location Service will be called. This call is accompanied by an intent that contains the current location as an object of the class Location.

```
public int onStartCommand(Intent intent, int flags, int startId) {
    location = intent.getParcelableExtra(FusedLocationProviderApi.KEY_LOCATION_CHANGED);
    locationChanged(location);
    return START_NOT_STICKY;
}
```

With these steps MobilitApp application receives updates about the user location changes.

### 3.1.2.2 Activity Recognition

The workflow followed to obtained activity updates is quite similar to the location ones. Many methods used for activity recognition tasks are implemented inside Location Service class; therefore, when activity updates take place there is a need to call this class. The explanation is done class by class and it starts at the initialization statement and ends with the activity updated.

#### STARTING ACTIVITY RECOGNITION SERVICE

Initialization process is the same as Location Service but instead of "LOC" the type selected is "AR". Requester is instantiated for Activity Recognition Service and then it will call Activity Recognition Service class, which is implemented as an *IntentService* [15] instead of Service. *IntentService* is handled in another thread and stops itself when finishes.

Just to remember the process the instance to Requester class and the call to Activity Recognition Service are wrote down below this paragraph.

When the incoming intent contains an update the new activity could be obtained with *getMostProbableActivity()* method and send it to Location Service class using *sendActivityLocationBroadcast(…)*.

```java
protected void onHandleIntent(Intent intent) {

  // If the incoming intent contains an update
  if (ActivityRecognitionResult.hasResult(intent)) {

    // Get the update
    ActivityRecognitionResult result = ActivityRecognitionResult.extractResult(intent);

    // Get the most probable activity
    DetectedActivity mostProbableActivity = result.getMostProbableActivity();

    // Get an integer describing the type of activity

    activityType = mostProbableActivity.getType();
    activityName = getNameFromType(activityType);

    /*
     * At this point, you have retrieved all the information for the
     * current update. You can display this information to the user in a
     * notification, or send it to an Activity or Service in a broadcast
     * Intent.
     */
    Intent intentActivity = new Intent ("activityRecognition");
    sendActivityLocationBroadcast(intentActivity, activityName);
  }
}
```

## ACTIVITY PROCESSING ALGORITHM

**LOCATION SERVICE CLASS**

The next part of the activity recognition algorithm is done in Location Service class. This class contains multiple methods to process the updated data.

First explained method the *Broadcast Receiver* [16] that obtains activity updates sent by the Activity Recognition Service every 5 seconds.

```java
private final BroadcastReceiver activityReceiver = new BroadcastReceiver() {

@Override
  public void onReceive(Context context, Intent intent) {

    activityProcessing(intent.getStringExtra("activity"));
  }
};
```

Inside Broadcast Receiver, the second method (*activityProcessing(…)*) is called with the activity name as argument. The activity name has been received as an *Extra* inside the *Intent*.

*activityProcessing(…)* counts the consecutive activity updates during the segment interval time. This method has a counter for each activity and every time an update is received the counter of the corresponding activity increases.

```java
private void activityProcessing(String _activity) {

    switch (_activity) {

    case "vehicle":
        vehicle_count++;
        break;
    case "bicycle":
        bicycle_count++;
        break;
    case "on_foot":
        on_foot_count++;
        break;
    case "still":
        still_count++;
        break;
    case "unknown":
        unknown_count++;
        break;
    }
}
```

The last method is called every 2 minutes, the segment interval time, and estimates which activity is done by the user. The arguments of *activityEstimation()* are the counters of the method explained before, and considering their values the algorithm decides which is the activity. The activity with more number of samples is the returned result, but if two or more activities have the same number of samples a priority order is set up. The order is the following: vehicle, bicycle, on_foot and still.

```java
private String activityEstimation(int _isOnfoot, int _isBicycle, int _isStill, int
_isVehicle, int _isUnknown) {

    ArrayList<Integer> max = new ArrayList<Integer>(5);

    max = findMax(_isOnfoot, _isBicycle, _isStill, _isVehicle, _isUnknown);

    switch (max.size()) {

    case 1:
        return (toActivity(max.get(0)));

    case 2:
        if (max.get(0) == 3 || max.get(1) == 3) {
            return toActivity(3);
        } else if (max.get(0) == 1 || max.get(1) == 1) {
            return toActivity(1);
        } else if (max.get(0) == 0 || max.get(1) == 0) {
            return toActivity(0);
        } else {
            return toActivity(2); //
        }
    case 3:
        if (max.get(0) == 3 || max.get(1) == 3 || max.get(2) == 3) {
            return toActivity(3);
        } else if (max.get(0) == 1 || max.get(1) == 1 || max.get(2) == 1) {
            return toActivity(1);
        } else {
            return toActivity(0);
        }
    case 4:
        if (max.get(0) == 3 || max.get(1) == 3 || max.get(2) == 3 || max.get(3) == 3) {
            return toActivity(3);
        } else {
            return toActivity(1);
        }
    case 5:
        return toActivity(2);
    }
    return "still";
}
```

*toActivity()* was developed to make interpretation easier when a result is returned. This method translates activity index established by Google Services to the name of the related activity.

This is the process that follows MobilitApp to receive updates about activity information.

## 3.2    Local Data and Modeling Transport´s Behaviour

The purpose of the present research project is to find an alternative method which allow detecting user´s activity and distinguish among the transports. After testing external services as the APIs mentioned before, the developer team has decided to study the way of obtaining data from mobile´s sensor and search for a transport modalities patterns.

The proposed solution is listening the accelerometer sensor; in the next section the sub-application that executes this task is explained. With this new functionality sample data from distinct means can be obtained and afterwards finding patterns for each one. These patterns will be the key in a future machine learning algorithm that would be implemented to substitute the transport activity detection described at the beginning of the chapter. The study which is going to be carried out is focused on motorized transports.

## 3.3   Listening Mobile Sensors

When the Accelerometer Listener was being developed the idea of obtaining data from more sensors appeared, then the developers decided to monitor the gyroscope as well because it is another sensor that could provide motion information. The later study is basically based on the study of acceleration patterns, but having data from another source could be helpful in a future.

### 3.3.1  Motion Sensors on Android devices. Accelerometer & Gyroscope

Android platform allow access to several sensors that monitor the motion of the mobile. Mobile sensors can be hardware-based, implemented physically on the device, or software-based, their data is derived from another sensors. Both sensors required, *Accelerometer* and *Gyroscope* [17], are hardware-based. One important point to pay attention when an application uses Android platform sensor is the power consumption. The next table shows average power consumption of different uses:

| Sensor/Feature | Average Power Consumption |
|---|---|
| Accelerometer(Hardware) | 0.23 mA |
| Magnetic Field(Hardware) | 6.8 mA |
| Orientation(Software) | 13.13 mA |
| Gyroscope(Hardware) | 6.1 mA |
| WiFi | 330 mA |
| 3G | 210 mA |

**Table 3-1: Mobile Components Consumption**

The monitored sensors have low power consumption. Comparing with other characteristics of the phone, sensors are not the main cause of dropping off the battery.

For accessing the sensor is necessary to add them as a feature used in the *Manifest* [18].

```
<uses-features
        android:name="android.hardware.sensor.accelerometer"
        android:required="true" />
<uses-features
        android:name="android.hardware.sensor.gyroscope"
        android:required="true" />
```

To conclude this explanation about the sensors in the Android platform an image of the coordinate system used by them is exposed.

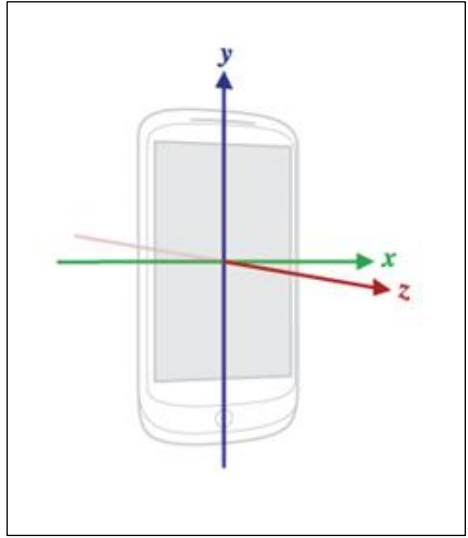

**Fig. 3-1: Axis Orientation**

## 3.3.2 Accelerometer Sensor Listener Development

Accelerometer Sensor Listener (ASL) has been developed as an independent application with the final purpose of implementing it inside MobilitApp. Its task is simple, reading data from mobile´s accelerometer, but this simplicity it is not a synonym of dispensable or inefficient. The ASL functionality is scalable, the same code with some modifications can be used to obtain data from other inner mobile sensors, and its potential and possibilities have been demonstrated in the current project. It has been a key element in the research project.

The ASL algorithm is explained in the lines below and most important parts of the code are shown to help reader´s comprehension. The description details which methods are implemented in each class to make clear the execution process.

### 3.3.2.1 Launching ASL

**SENSOR DATA CLASS**

An object for accelerometer´s data has been created. It consists of four variables; *timestamp* for store the moment when a change in acceleration takes place and *x*, *y* and *z* to store each acceleration's change in the corresponding axis. Four methods have been also implemented to retrieve the value of each variable from anywhere of the code.

```java
public class SensorData {

    private String timestamp;
    private double x;
    private double y;
    private double z;

    public SensorData(String timestamp, double x, double y, double z){
        this.timestamp = timestamp;
        this.x = x;
        this.y = y;
        this.z = z;
    }
    public String getTimestamp(){
        return timestamp;
    }

    public double getX() {
        return x;
    }

    public double getY() {
        return y;
    }

    public double getZ() {
        return z;
    }

}
```

**START CLASS**

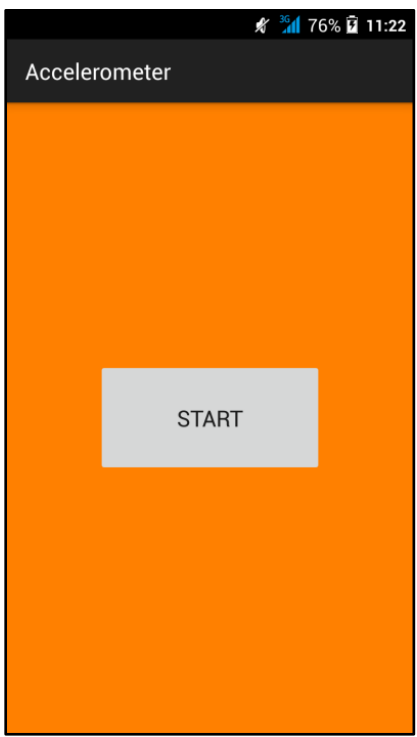

Fig. 3-2: Launching ASL

Start Class is the *Launcher Activity Class*, the *Launcher Activity* is the one that appears firstly when an application is opened. In this case is an orange screen with a START button which takes the user to the main activity, Data Screen. The orange colour has been chosen for the background because it is MobilitApp´s logo colour, Figure 3-2 illustrates it.

When the user clicks on the START button Data Screen is called after that, *Intent* [19] is used to call the main *Activity* [20].

```java
public void onClick(View view){
    // Start Button
    //When we call the Main activity we send a boolean message with the Intent
    Intent i = new Intent(this,DataScreen.class);
    // onStartMessage helps us to know if the action is started or not
    i.putExtra("onStartMessage",true);
    // the activity starts
    startActivity(i);
}
```

### 3.3.2.2    Data Screen

**DATA SCREEN CLASS**

This activity contains the largest part of the Accelerometer Sensor Listener code. For making the explanation more understandable it will be divided in many parts, in each of these parts a functionality of the algorithm is described.

### 3.3.2.3    Interface and Interaction with the user

To visualize the data read from the accelerometer a layout like the Figure 3-3 has been designed.

The data is shown in a four columns table, one column for timestamp value and the other three for the acceleration data, one for each axis. Start, Stop and Exit buttons allow the user to interact with the application lifecycle.

Over these buttons there is a *timer* which will be described in detail in a next section. On the *onCreate()* method the timer is set to zero.

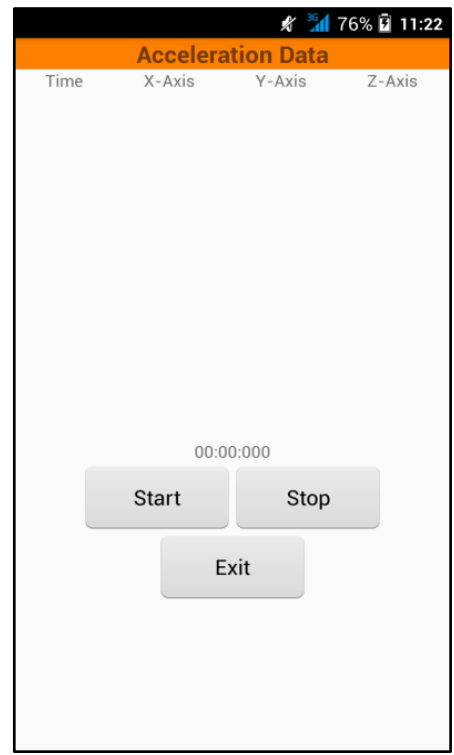

**Fig. 3-3: Accelerometer Listener**

When the Start button is clicked Accelerometer and Gyroscope sensors are instantiated, remember that data from gyroscope is also acquired. *Sensor Manager* [21] is used to make actions over the sensors, *getSystemService()* statement provides credentials to this variable for accessing the sensors. With *getDefaultSensor()* Accelerometer and Gyroscope are instantiate to variables and *registerListener()* is the method which creates an *Event Listener* [22] for a defined sensor at the given frequency. Every time that acceleration changes the Listener calls *onSensorChanged()* where all the sensor´s information is processed.

```
private SensorManager mSensorManager;
private Sensor mAccelerometer;
private Sensor mGyroscope;
...
public void onStartClick(View view) {
// Instance the accelerometer and gyroscope
mSensorManager = (SensorManager) getSystemService(Context.SENSOR_SERVICE);
mAccelerometer = mSensorManager.getDefaultSensor(Sensor.TYPE_ACCELEROMETER);
mGyroscope = mSensorManager.getDefaultSensor(Sensor.TYPE_GYROSCOPE);
...
mSensorManager.registerListener(this, mAccelerometer, SensorManager.SENSOR_DELAY_NORMAL);
mSensorManager.registerListener(this, mGyroscope, SensorManager.SENSOR_DELAY_NORMAL);
...
```

### 3.3.2.4    Timer Thread

A parallel thread has been created to execute the timer task. There are some ways to implement new threads in an Android activity; in this case the *Handler* [23] class and the interface *Runnable* [24] are combined to build the function. The reason of creating a new execution thread for the timer function is to not interfere in the main activity of application, listening accelerometer and gyroscope sensors.

In the moment the user decides to start listening sensors´ data the current time is kept, *currentTimeMillis()*, and then a call to the *Runnable interface*, which executes the timer procedure, is generated. These actions are done in the same statement where the accelerometer and gyroscope listeners have been created.

```
// Moment in which the accelerometer is going to store data
start_time = System.currentTimeMillis();

//call the run method of timerRunnable
timerHandler.post(timerRunnable);
```

*Handler* use is to perform the timer *Runnable* in a different thread. *Runnable* just implements one method, *run()*, where remains all the time code. The start timestamp registered before is compared with the current time to obtain how long is the app listening sensors and finally show it to the user in an understandable format. The last statement calls the *run()* method again and so on and so forth, the importance of this line is remarkable because without it the timer would not work correctly.

```
Handler timerHandler = new Handler();
Runnable timerRunnable = new Runnable() {
@Override
public void run() {
//this method emulates a timer

    currentTime = System.currentTimeMillis() - start_time;

    TextView timerText = (TextView) findViewById(R.id.timer_text);
    timerText.setText(time(currentTime, "mm:ss:SSS"));

    //Very Important! this Handler method calls the runnable again
    timerHandler.post(this);
    }
};
```

A function to show the time value in the desired format has been implemented.

```
//Gives format to a date
//Creates a format class with a specific format pattern
public static String time (long milliSeconds, String dateFormat) {

    SimpleDateFormat formatter = new SimpleDateFormat(dateFormat);
    //Get the current date and time
    Calendar calendar = Calendar.getInstance();
    //Converts the time into milliseconds
    calendar.setTimeInMillis(milliSeconds);
    String time = formatter.format(calendar.getTime()).toString();
    return time;
}
```

When the Stop button is pushed the timer stops and the Runnable is not call anymore.

```
timerHandler.removeCallbacks(timerRunnable);
```

### 3.3.2.5    Collecting Sensor Values

Once registered the accelerometer and the gyroscope the method *onSensorChanged()* is ready to register every change in the sensors' values. The data will be stored in Sensor Data arrays, remember that Sensor Data is the object developed to save read values on it.

```
//Specify the object type which will be stored in the ArrayList
private ArrayList<SensorData> AccelerationData = new ArrayList<SensorData> ();
//Specify the object type which will be stored in the ArrayList
private ArrayList<SensorData> GyroscopeData = new ArrayList<SensorData> ();
```

First of all a sensor classification is done, to store the information in its corresponding array.

```
...
if (event.sensor.getType() == Sensor.TYPE_ACCELEROMETER){
...
} else if (event.sensor.getType() == Sensor.TYPE_GYROSCOPE){
...
}
```

Then another condition has to be accomplished. The variation in the acceleration or angular rate has to be greater than a threshold. This threshold defines from which value the data could be considered good for the movement pattern study.

The **acceleration threshold** has been established in the following way:

Acceleration equation:

$$acceleration = \frac{\Delta velocity}{\Delta time}$$

- Which velocity variation is desired to be detected? Velocity variation is the difference between final velocity and initial velocity.

  Some values have been tested and the best behaviour is achieved with 4m/s.

- The collecting data window is 20 seconds, then a value between 0 and 20 will be the selected one.10 seconds has been chosen as the time variation value, it is lower than the window´s value but both are in the same order of magnitude.

$$Threshold = \frac{4 \, m/s}{10 \, s} = 0,4 \, m/s^2$$

The **angular rate** threshold has been established in a similar way:

Angular rate equation:

$$angular \, rate = \frac{\Delta angle}{\Delta time}$$

- Which angle variation is desired to be detected? Angle variation is the difference between final angle and initial angle.

   Some values have been tested and the best behaviour is achieved with 30º, *0, 5 radians* approximately.

- Time variation is not the same than in acceleration threshold because quick variations are desired, and for this reason 1 second has been established as time difference.

$$Threshold = \frac{0,5 \, rad}{1 \, s} = 0,5 \, rad/s$$

Before storing the data it is necessary to process it. The mobile engine´s vibrations cause wrong sensor readings. To avoid this issue a *Low Pass Filter* is applied to the input data because it can omit the high frequency noise added to the monitored signal.

The Low Pass Filter implemented looks as follows:

$$y[i] = y[i-1] + \alpha * (x[i] - y[i-1])$$

If it is developed the final expression is:

$$y[i] = \alpha * x[i] + (1 - \alpha) * y[i-1]$$

Where x[i]is the current input, y[i] is the current output, y[i-1] is the previous output and α is the low pass filter constant, also called cut-off constant. The filter constant has been calculated with the next relation:

$$\alpha = \frac{t}{t + dT}$$

Being t latency added by the filter which we use to fix the listening period, and dT the event delivery rate established when the sensor listener was registered, SENSOR_DELAY_NORMAL tag value (200μs) in this case.

```
x = alpha*event.values[0] + (1-alpha)*AccelerationData.get(position - 1).getX();
y = alpha*event.values[1] + (1-alpha)*AccelerationData.get(position - 1).getY();
z = alpha*event.values[2] + (1-alpha)*AccelerationData.get(position - 1).getZ();
```

After processing sensor´s data, they are stored in a Sensor Data object and then in the arrays. The Figure 3-4 shows how data is represented in the screen.

```
SensorData data_accelerometer = new SensorData(timestamp, x, y, z);
AccelerationData.add(data_accelerometer);
```

Fig. 3-4: Accelerometer Data

### 3.3.2.6    Saving Samples in a File

All the collected data is saved in a CSV file when the Stop button is pushed, a file for each sensor is created. *onStopClick()* calls DataStorage class to store the data in the mobile, the files are kept until they are uploaded to the server.

```
public void onStopClick(View view) {

//Unregister the Listener for all the sensors
mSensorManager.unregisterListener(this);
//Stops timerRunnable
timerHandler.removeCallbacks(timerRunnable);

new DataStorage().execute(AccelerationData);

new DataStorage().execute(GyroscopeData);
}
```

DataStorage is an *AsyncTask* [25] that receive the Sensor Data array as parameter. *AsynTask* is another method to execute procedures in different threads.

Sensor Data implemented methods now are used to pass collected values to new variables. *writeNext()* writes variable values in the file created.

```java
for (int i = 0; i < length; i++) {

    data = params[0].get(i);

    timestamp = data.getTimestamp();
    x = data.getX();
    y = data.getY();
    z = data.getZ();

    //Only 3 first decimals are stored
    writer.writeNext(new String[]{timestamp, String.format("%.3f", x), String.format("%.3f",
     y), String.format("%.3f", z)});
}
```

The storing data process will be ended by uploading this data to MobilitApp server, to do that the next step is calling the class built for this task. UploadFile implements *Runnable* interface, so it can be used as an argument for a Thread object. An UploadFile object is created and then passed it as the argument of a Thread class, a parallel thread is needed because this task must be done in background. *Thread* [26] constructs a new execution process with the code of the provided object.

```java
UploadFile filetoupload = new UploadFile(csv_file.toString());

Thread UploadFile = new Thread(filetoupload);

UploadFile.start();
```

The state of the Upload File process is controlled because once the collected data will be in the server, it will be deleted from mobile´s memory.

```java
// Get the state of the UploadFile thread and when it´s finished then delete the sensor
data from the mobile
while(UploadFile.getState() != Thread.State.TERMINATED){

}
//Delete sensor files
File delete_file = new File(csv_file.toString());
Boolean delete = delete_file.delete();
```

### 3.3.2.7    Upload Data File

To transfer the files to a server a connection is needed, so an HTTP connection is started. HttpURLConnection Android class allows connecting to a concrete URL. Once the contact has been established, the request command method will be sent to the server. *POST* is the mode used to upload the data and to send it an *outputStream* has been built over the connection.

```java
HttpURLConnection connection = null;
DataOutputStream outputStream = null;

String urlServer = "http://mobilitapp.noip.me/csv/post_date_receiver.php";

    ...

    FileInputStream fileInputStream = new FileInputStream(new File(filePath));

    URL url = new URL(urlServer);
    connection = (HttpURLConnection) url.openConnection();

    // Set HTTP method to POST.
    connection.setRequestMethod("POST");

    outputStream = new DataOutputStream( connection.getOutputStream() );

    ...

    bytesRead = fileInputStream.read(buffer, 0, bufferSize);

    while (bytesRead > 0)
    {
        outputStream.write(buffer, 0, bufferSize);
        bytesAvailable = fileInputStream.available();
        bufferSize = Math.min(bytesAvailable, maxBufferSize);
        bytesRead = fileInputStream.read(buffer, 0, bufferSize);

    }
```

On the server side there is a PHP script which is responsible for receiving data. Take special attention to the URL address, its value is the folder where the previous script is located. In fact the application (client) is asking for *posting* data on the server, and the PHP code is answering to client´s request and it is executing the storage operation. Each day a folder is created and the received files are archived according to their date.

```php
<?php

$hoy = date("j-m-Y");
$directory = $hoy."/";
$target_path = "/home/pi/web/csv/";
$target_path = $target_path . basename( $_FILES['uploadedfile']['name']);
if(move_uploaded_file($_FILES['uploadedfile']['tmp_name'], $target_path))
{
    echo "The file ".  basename( $_FILES['uploadedfile']['name']).
 " has been uploaded";
}
else
{
    echo "There was an error uploading the file, please try again!";
}
?>;
```

# 4. Upgrading MobilitApp

## 4.1 Including the Accelerometer Sensor Listener

In this chapter is explained how the previous development is incorporated to the MobilitApp algorithm. Adding ASL classes and changing the current code and the workflow are the essential modifications on the application.

The next two sections describe two components which were not part of the Acceleration Sensor Listener design but now are necessary for the correct behaviour of the application. Later, the new execution process is detailed pointing on how the ASL components interact with MobilitApp current algorithm.

### 4.1.1 Sensor Data and Sensor Data List

Sensor Data is the object created to store the listened values and it will be included in the final design. The data will be passed from the Accelerometer Sensor Listener to the MobilitApp code and to reach this a modification on the Sensor Data class is required. Implementing *Parcelable* [27] interface helps to pass data between the application's components.

**SENSOR DATA LIST**

The Sensor Data List class is created to keep the data in an array of Sensor Data objects. When the collected sensor values are sent to another activity the method used is *writeToParcel()*, this one gets the values of each component along the *ArrayList* [28] and writes into a *Parcel*.

```java
public class SensorDataList extends ArrayList<SensorData> implements Parcelable {
...
@Override
public void writeToParcel(Parcel dest, int flags) {

    int size = this.size();

    dest.writeInt(size);

    for (int i = 0; i < size; i++) {

        SensorData data = this.get(i);

        dest.writeString(data.getTimestamp());

        dest.writeDouble(data.getX());

        dest.writeDouble(data.getY());

        dest.writeDouble(data.getZ());

        }
    }
```

The values written by the *writeToParcel()* method are recovered from the *Parcel* and stored in a new Sensor Data List built in the destination class. CREATOR is an interface that generates instances of the *Parcelable* implemented class from the *Parcel*.

```java
public static final Creator<SensorDataList> CREATOR = new Creator<SensorDataList>() {
    @Override
    public SensorDataList createFromParcel(Parcel in) {
        return new SensorDataList(in);
    }

    @Override
    public SensorDataList[] newArray(int size) {
        return new SensorDataList[size];
    }
};
```

The data array is scanned in order and in every iteration a Sensor Data is created to store the information from this position on the list. Summarizing, the information collected is restored from a *Parcel* [29] once the object has been transferee.

```java
private void readFromParcel(Parcel in) {

    this.clear();

    //First we have to read the list size

    int size = in.readInt();

    //Reading SensorData attributes

    //Order is fundamental

    for (int i = 0; i < size; i++) {

        SensorData data = new SensorData();

        data.setTimestamp(in.readString());

        data.setX(in.readDouble());

        data.setY(in.readDouble());

        data.setZ(in.readDouble());

        this.add(data);

    }

}
```

Now, with this new implementation transferring data between application´s components is possible. Particularly, this new functionality will be used when the collected data is sent to the class which saves the sensor values into a file.

## 4.1.2  Select Transport Fragment

In the collecting data process user´s help is needed, to know which transport modality they are using a vehicle selection list will be handling this function. The developer team will analyze the mobility data to find patterns which define each mean, and knowing the origin of each group of values will be very helpful.

First of all the application ask the user if they are agree to contribute providing to MobilitApp his mobility information, then a selectable list with transportation modes is displayed to choose which one is going to be used. The final study is focused on the most important transportation means in the metropolitan area of Barcelona; Car, Motorbike, Bus, Metro and Train. The two screens shown below appear when the user initializes MobilitApp.

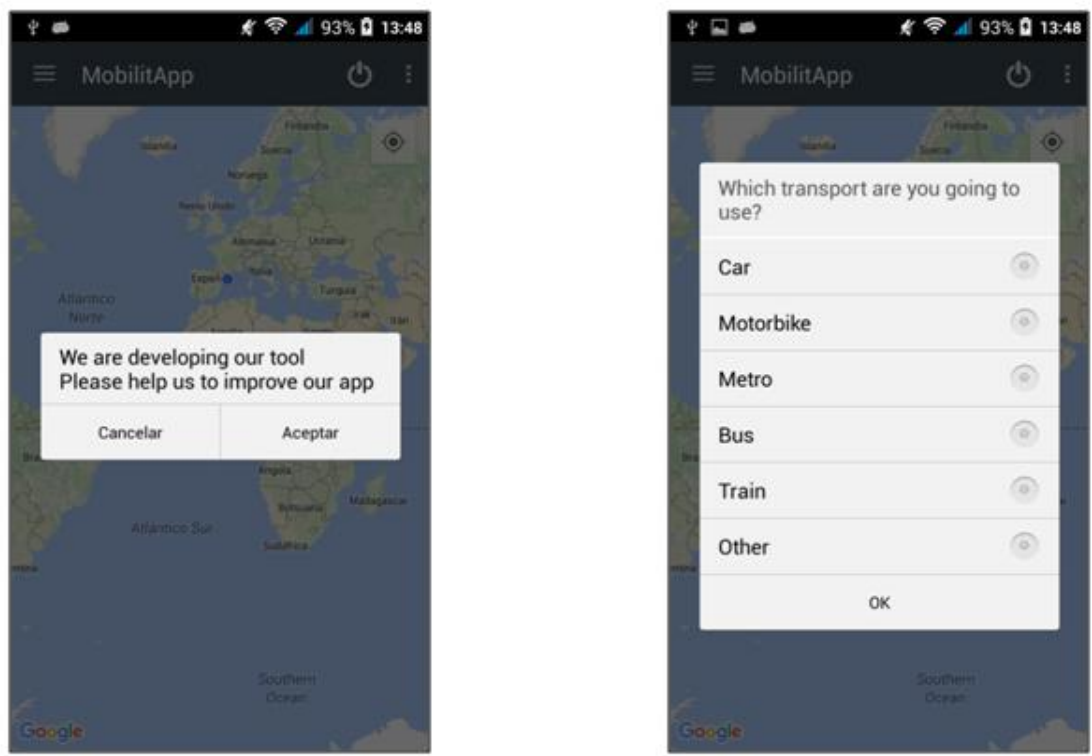

**Fig. 4-1: Select Transport Fragment**

**SELECT TRANSPORT FRAGMENT**

To deploy this feature the *DialogFragment* [30] class is used, it displays a dialog window floating on top of its activity´s window. User´s transport selection will be saved in a *static* variable (*public static String selection*) to keep this value during all the application lifetime cycle and to allow accessing it from each part of the code.

In following sections Select Transport Fragment procedure is clarified, where it takes place inside MobilitApp execution process and where is the selected transport´s name retrieved.

```java
public class SelectTransportFragment extends DialogFragment {

String [] mtransports = {"Car","Motorbike","Metro","Bus","Train","Other"};
public static String selection;

public Dialog onCreateDialog(Bundle savedInstanceState) {

    AlertDialog.Builder builder = new AlertDialog.Builder(getActivity());

    builder.setTitle("Which transport are you going to use?")
        .setSingleChoiceItems(mtransports, -1, new DialogInterface.OnClickListener() {

            public void onClick(DialogInterface dialog, int which) {
                switch (which){
                    case 0:
                        selection = mtransports[which];

                        break;
                    case 1:
                        selection = mtransports[which];

                        break;
                    case 2:
                        selection = mtransports[which];

                        break;
                    case 3:
                        selection = mtransports[which];

                        break;
                    case 4:
                        selection = mtransports[which];

                        break;
                    case 5:
                        selection = mtransports[which];

                        break;
                }

            }
        }).setPositiveButton("OK",new DialogInterface.OnClickListener(){

        public void onClick(DialogInterface dialog, int which) {
            Toast.makeText(getActivity(), selection, Toast.LENGTH_SHORT).show();
        }
    });

    return builder.create();
}
```

# 4.2 Accelerometer Sensor Listener Running in Background inside MobilitApp

MobilitApp function is tracking citizen´s daily routes and since now another service is going to run on background. The Accelerometer Sensor Listener subapplication is implemented inside MobilitApp and both services coexist in the same context. The next entries specify how ASL has been integrated on MobilitApp workflow. The description is done by classes and in order of execution.

## 4.2.1 Initializing MobilitApp

Just to remember, when the application starts it requests Location and Activity Recognition updates, then intent for each service is created. The Location information is updated every 20 seconds and the Activity every 5 seconds.

**MAIN ACTIVITY**

Now when MobilitApp is initialized, a selectable transport list is displayed, at the same time the Location and Activity Recognition updates requests are executed. The image below shows how the *AlertDialog* [31] instantiates the SelectTranportFragment.

```java
AlertDialog.Builder alertDialogBuilder = new AlertDialog.Builder(this);
alertDialogBuilder.setMessage("We are developing our tool Please help us to improve our app")
.setPositiveButton(android.R.string.yes, new DialogInterface.OnClickListener() {
@Override
public void onClick(DialogInterface dialog, int which) {

    SelectTransportFragment selectTransportFragment = new SelectTransportFragment();
    selectTransportFragment.show(getSupportFragmentManager(),"select_transport");
    }
})
.setNegativeButton(android.R.string.no,new DialogInterface.OnClickListener(){
@Override
public void onClick(DialogInterface dialog, int which) {
    dialog.cancel();
    }
})
.show();
```

## 4.2.2 Calling Accelerometer Sensor Listener

**LOCATION SERVICE**

After 2 minutes the Activity Recognition updates are processed and an estimation of the transport used is run. Once the estimation has concluded the ASL algorithm service is started, except if the estimated activity is *still* or *on_foot*. Remind that the new solution is focused on motorized transports.

```java
// At this point depending on the API results we will call a subapp that collects
accelerometer/gyroscope data or won't

activityName = activityEstimation(on_foot_count, bicycle_count, still_count,
vehicle_count,
    unknown_count);

if(activityName != "still" ||  activityName != "on_foot") {
    Intent Sensor_Listener = new Intent(this, Sensor_Listener.class);

    startService(Sensor_Listener);
}
```

### 4.2.2.1    Priority on Estimation Process

The priority order algorithm has been changed, vehicle continues having the maximum priority but now it have more presence. Even vehicle would not be the most received activity, if the difference between vehicle and the highest activity counter is lower than three the estimated activity is vehicle.

```java
private String activityEstimation(int _isOnfoot, int _isBicycle, int _isStill, int _isVehicle,
int _isUnknown) {

    ArrayList<Integer> max = new ArrayList<Integer>(5);

    max = findMax(_isOnfoot, _isBicycle, _isStill, _isVehicle, _isUnknown);

    switch (max.size()) {

    case 1:
        if (max.get(0) == 3)
            return (toActivity(max.get(0)));

        else {
        if ((max.get(0) == 1)&&(_isBicycle - _isVehicle) > 3) {

            return toActivity(1);
            }
        else if ((max.get(0) == 0)&&(_isOnfoot - _isVehicle) > 3) {
            return toActivity(0);
        }
        else if ((max.get(0) == 2)&&(_isStill - _isVehicle) > 3) {
            return toActivity(2);
        }
        else {
            return toActivity(3);
            }
        }
```

```
case 2:
    if (max.get(0) == 3 || max.get(1) == 3) {
        return toActivity(3);
    } else if ((max.get(0) == 1 || max.get(1) == 1)&&(_isBicycle - _isVehicle > 3)) {

        return toActivity(1);
    } else if ((max.get(0) == 0 || max.get(1) == 0)&&(_isOnfoot - _isVehicle > 3)) {

        return toActivity(0);
    } else if (_isStill - _isVehicle > 3){
        return toActivity(2);

    } else {
        return toActivity(3);
    }
case 3:
    if (max.get(0) == 3 || max.get(1) == 3 || max.get(2) == 3) {
        return toActivity(3);
    } else if ((max.get(0) == 1 || max.get(1) == 1 || max.get(2) == 1)&&(_isBicycle - _isVehicle
    > 3)) {
        return toActivity(1);
    } else if (_isOnfoot - _isVehicle > 3){
        return toActivity(0);

    } else {
        return toActivity(3);
    }

    }
```

```
    case 4:
        if (max.get(0) == 3 || max.get(1) == 3 || max.get(2) == 3 || max.get(3) == 3) {
            return toActivity(3);
        } else if(_isBicycle - _isVehicle > 3){
            return toActivity(1);
        } else {
            return toActivity(3);
        }
    case 5:
        return toActivity(2);
    }

    return "still";

}
```

### 4.2.3 Sensor Listener Intent Service

**SENSOR LISTENER CLASS**

Every time the Sensor Listener is called, the *onHandleIntent()* method is in charge of registering and monitoring the sensors. But the accelerometer will not be listened during all the application lifecycle, just during a period of time every 2 minutes. Firstly, Sensor Listener service and Location Service have to be synchronized, the first one waits until the other wakes it up with new data.

```java
protected void onHandleIntent(Intent intent) {
    ...
    //Stop and wait until 20 seconds would have passed
    synchronized (send_data){
        try {
            Log.v(MyTAG,"Wait for Data");
            send_data.wait();
        } catch (InterruptedException e) {
            e.printStackTrace();
        }
    }
    ...
}
```

On the Timer Thread code 20 seconds is fixed as the time interval while the ASL is active, and consequently, sensor data is received. The window duration is established on the timer emulator because it is the class which has been created to control the monitoring time.

```java
Runnable timerRunnable = new Runnable() {
    @Override
    public void run() {
    //this method emulates a timer

    currentTime = System.currentTimeMillis() - start_time;

    //Very Important! this Handler method calls the runnable again
    timerHandler.post(this);

    if(currentTime > 20000){

    synchronized (send_data){
        send_data.notify();
        }
    }

    }
};
```

*send_data.wait()* freeze the thread until another thread, *Timer Thread*, would call *send_data.notify()*. The locked thread is now able to continue. To sum up, accelerometer and gyroscope are listened 20 seconds every 2 minutes inside MobilitApp workflow.

## 4.2.4 Data Storage and Uploading Files to Server

Once the collection data process has finished, the values kept in the Sensor Data List will be stored in a file. As it was mentioned before the sensor arrays will be transferred between ASL algorithm and MobilitApp code. Here the importance of the *Parcelable* interface is highlighted.

```
new LocationService().StoreDataSensor(AccelerationData);

new LocationService().StoreDataSensor(GyroscopeData);
```

The method introduced in the next image is implemented on Location Service class. Its functionality is passing the SensorDataList to the class which stores the data in a file.

```
public void StoreDataSensor(SensorDataList sensorDataList) {

new DataStorage(this).execute(sensorDataList);

}
```

**DATA STORAGE CLASS**

This class behaves in the same way as the explained in ASL design, *3.3.2.6 - Saving Samples in a File*, with the added aspect that the file´s name includes the selected transport´s name. The static variable seen before (*selection*) is now instantiated to retrieve its value. In this way, when the files are going to be uploaded to the server, they include the transport modality name and the date when the activity happened.

```
//at this point the selected transport at the beginning of the service
//is retrieved and used in the file's name

transport_selected = SelectTransportFragment.selection;

if(sensorDataList.getSensorName().equals("AccelerationData")){
    filename.append("acceleration_"+transport_selected+"_"+date+".csv");
}

if(sensorDataList.getSensorName().equals("GyroscopeData")) {
    filename.append("gyroscope_"+transport_selected+"_"+date+".csv");
}
```

**UPLOAD FILE CLASS**

Upload File class is equivalent of the one which has been described in *3.3.2.7-Upload Data File*, and the procedure in relation with the connection server is still the same.

## 4.3    Technological Infrastructure

To complete this chapter the current section gives an overview about which kind of technology is involved on MobilitApp project.

**Platform**

MobilitApp is developed over **Android platform** which allows developers using **Google APIs** to add different features to the apps. It uses Maps, Places, Directions, Location Services and Activity Recognition.

**Open Data**

MobilitApp integrates **real-time traffic** state information provided by the Barcelona City Hall, **Open Data BCN – eGovernment (*http://opendata.bcn.cat/opendata/ca*)**, and the **traffic incidences**, **DGT – eTraffic (http://infocar.dgt.es/etraffic/)**.

**Server**

All the collected data is sent to a server in *csv* format file. The server is assembled over a Raspberry Pi device, Annex A, and it provides the **web** service, the **database** service for storing users´ profile and the tracking citizens´ activity, and accumulates all the **sensor data files**.

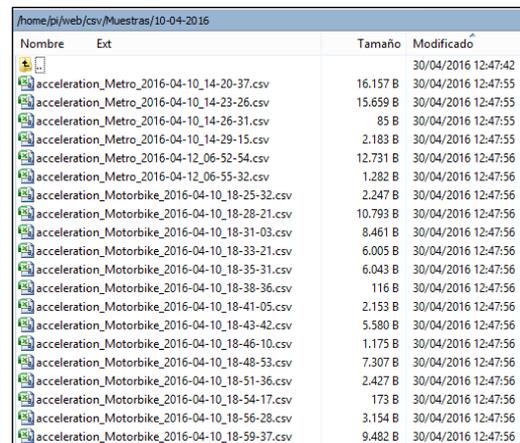

Fig. 4-2: Sensor Data Files in Server

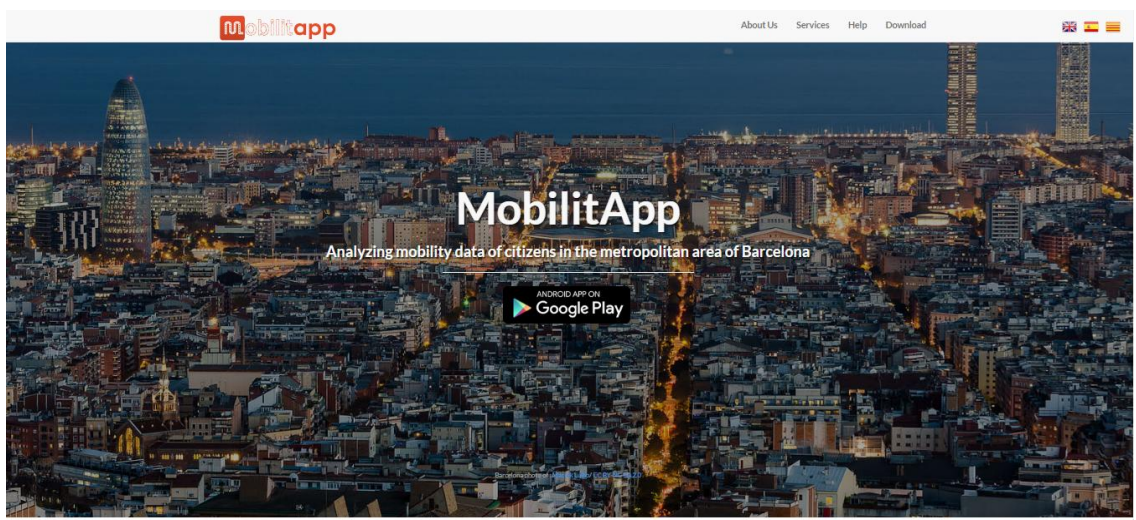

Fig. 4-3: MobilitApp Web Site

**Google Play Store**

MobilitApp has been uploaded to Play Store to make the installation easier and to reach a larger number of users. In addition when a modification on the application code is done, this can easily be upgraded.

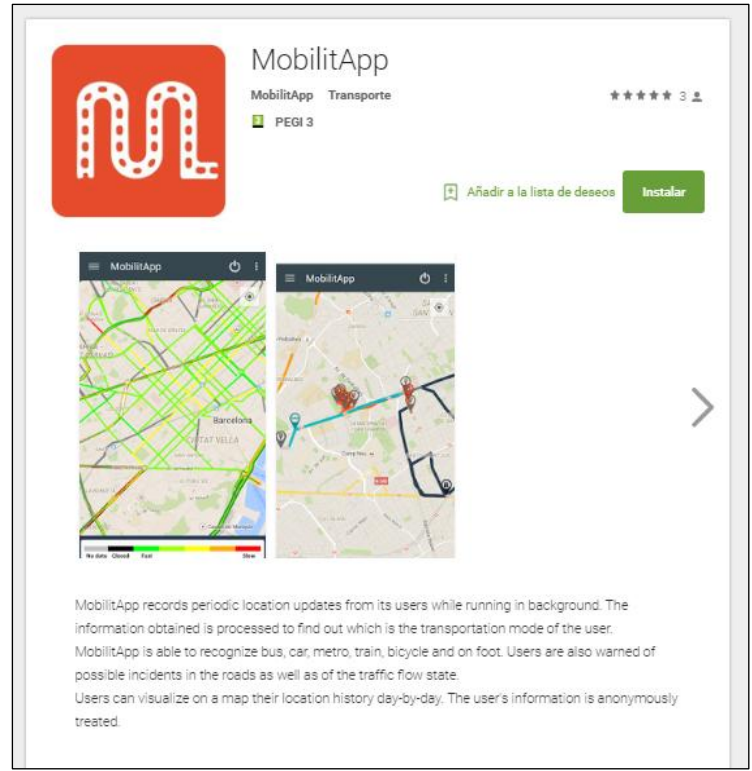

**Fig. 4-4: MobilitApp on Google Play**

# 5. Analyzing Mobility Data

In this chapter the final process of the research project is explained, the analysis of all the obtained information. With analysis´ results a final conclusion is going to be done, based on the premise that the different transport modalities have a different behaviours.

The explanation is done step by step, opening with the collecting process, passing through the corresponding analysis and finishing with a proposed solution.

## 5.1 Collecting Data

### 5.1.1 Description of the Context

The data has been collected between March 10[th] and April 15[th] 2016 with the help of MobilitApp users and the other members of the project team. The majority of the saved files from each person have the same time characteristic; the user opens MobilitApp in his daily commutes, so the time interval when the app is used is always the same. A larger number of users is needed to cover all the time zones and more possible journeys across the city.

Another issue that is necessary to comment is the mobile phone´s location during the activities. Statistically, people keep their mobile phone in their pocket or in a bag or handbag. In addition to these, when somebody travels by car he usually leaves his mobile phone on the dashboard. Then, the three main sites where people are used to carry his mobile are pocket, bag or handbag and dashboard´s car.

At the end of the collecting process 1.15 MB of sensor samples has been accumulated, stored at the server and organized by dates and transport mode. All the data has been collected from journeys inside the urban area of Barcelona except the train samples which has been obtained from a Barcelona – Cartagena round trip.

## 5.2    Analyzing Collected Data

Accelerometer data has been analysis with the purpose of finding patterns that allow activity recognition. In the current test the gyroscope information has not taken part in it, the sensor files still stored on the server for future researches.

Two analysis algorithms have been developed using the mathematical tool *MATLAB* [32]; a **Statistical Analysis of the Sampled Data** and a **Representation of the Horizontal Acceleration Magnitude**.

### 5.2.1  Statistical Analysis of the Sampled Data

In this study the statistical parameters of the data have been evaluated. In some samples the gravity component is not well estimated, then the longitudinal component (Horizontal Acceleration) could not being found. The Horizontal Acceleration is the measure which represents better the behaviour along the movement direction, because of that the analysis is done over this component.

The features considered are:  Mean, Standard Deviation, Variance, Minimum and Maximum Value, Range and Root Mean Square Error. In this case RMSE is the same as the STD because it has been calculate around the Mean.

The following figures show how was measured this statistical parameters. The code below opens the *csv* files where the data is stored and extract it into variables.

```matlab
% Open data file
fid = fopen('Muestras\Bus\acceleration_Bus_2016-04-20_19-51-27.csv');

% Open a file to write the parameters on it

fid2 = fopen('Bus_Analysis_2016-04-20_3.txt','wt');

% Read data in from CSV file

readData = textscan(fid,'%s %s %s %s','HeaderLines',1,'Delimiter',';');

% Remove double quotes from timestamp and z
readData{1,1}(:,1) = strrep(readData{1,1}(:,1),'"','');
readData{1,4}(:,1) = strrep(readData{1,4}(:,1),'"','');

fclose(fid);

% Extract data from readData
timestamp = readData{1,1}(:,1);
x = readData{1,2}(:,1);
y = readData{1,3}(:,1);
z = readData{1,4}(:,1);

% Convert timestamp into a date number
timestamp = datenum(timestamp,'MM:SS:FFF');

% Convert x,y and z cell arrays to double vectors
x_acceleration = str2double(x);
y_acceleration = str2double(y);
z_acceleration = str2double(z);
```

Having the data ready for being processed, the first features are calculated; **Mean**, **Variance** and **Standard Deviation**. Before, the acceleration is transformed into its absolute value; it is desired to obtain these parameters without regarding if they are positive or negative, just considering the magnitude value.

```matlab
% Calculate the mean, variance and standard deviation
% of the acceleration magnitude over each axis

x_abs = sqrt(x_acceleration.*x_acceleration);
y_abs = sqrt(y_acceleration.*y_acceleration);
z_abs = sqrt(z_acceleration.*z_acceleration);

% axis acceleration

mean_x = mean(x_abs);
variance_x = var(x_abs);
std_x = std(x_abs);
fprintf('\nx_mean = %f\n', mean_x );
fprintf('variance_x = %f\n', variance_x );
fprintf('x-standard deviation = %f\n', std_x );

mean_y = mean(y_abs);
variance_y = var(y_abs);
std_y = std(y_abs);
fprintf('\ny_mean = %f\n', mean_y );
fprintf('variance_y = %f\n', variance_x );
fprintf('y-standard deviation  = %f\n', std_y );

mean_z = mean(z_abs);
variance_z = var(z_abs);
std_z = std(z_abs);
fprintf('\nz_mean = %f\n', mean_z );
fprintf('variance_z = %f\n', variance_x );
fprintf('z-standard deviation  = %f\n', std_z );
```

**Maximum** and **Minimum** values are obtained to show the acceleration range for each mean.

```matlab
% Maximum and minimum values
fprintf('\nMaximum and Minimum coefficients.\n');

x_max_coeff = max(x_abs);
x_min_coeff = min(x_abs);
fprintf('\nx_max_coeff = %f\n', x_max_coeff );
fprintf('x_min_coeff = %f\n', x_min_coeff );

y_max_coeff = max(y_abs);
y_min_coeff = min(y_abs);
fprintf('\ny_max_coeff = %f\n', y_max_coeff );
fprintf('y_min_coeff = %f\n', y_min_coeff );

z_max_coeff = max(z_abs);
z_min_coeff = min(z_abs);
fprintf('\nz_max_coeff = %f\n', z_max_coeff );
fprintf('z_min_coeff = %f\n', z_min_coeff );

%%% Calcute Range %%%

fprintf('\nAcceleration Ranges\n');
fprintf('\nHorizontal Range = [%f,%f]\n',min(x_abs),max(x_abs));
fprintf('\nVertical Range = [%f,%f]\n',min(y_abs),max(y_abs));
fprintf('\nPosible Vertical Range = [%f,%f]\n',min(z_abs),max(z_abs));
```

The **Root Mean Square Error** operation requires the real mean value; taking the sign into account.

```matlab
%%% Root Mean Square Error %%%

mean_x = mean(x_acceleration);
mean_y = mean(y_acceleration);
mean_z = mean(z_acceleration);
...

    % RMS Error
RMS_Error_X = sqrt(sum((mean_x - x_acceleration).^2)/length(x_acceleration));
RMS_Error_Y = sqrt(sum((mean_y - y_acceleration).^2)/length(y_acceleration));
RMS_Error_Z = sqrt(sum((mean_z - z_acceleration).^2)/length(z_acceleration));
...

residual_x = abs(mean_x - x_acceleration);
x_samples = 0;

for k=1:length(residual_x)

    if(residual_x(k)<RMS_Error_X)

    x_samples = x_samples + 1;
    end

end

residual_y = abs(mean_y - y_acceleration);
y_samples = 0;

for i=1:length(residual_y)

    if(residual_y(i)<RMS_Error_Y)

    y_samples = y_samples + 1;
    end

end

residual_z = abs(mean_z - z_acceleration);
z_samples = 0;

for j=1:length(residual_z)

    if(residual_z(j)<RMS_Error_Z)

    z_samples = z_samples + 1;
    end

end
```

Finally a graphic result is added to make the results more understandable. **Histogram** [33] sorts the samples by values. Obtaining a visual result could provide a survey of how acceleration is distributed and why the statistics results have concrete values.

```
%%%% Plot Histogram %%%%

% A distribution of all the axis acceleration values is shown
% The component with more samples around 9.8 m/s2 will be considered
% gravity component

f2 = figure(2);
set(f2,'name','Histogram_Car_2016-03-10','numbertitle','off');
subplot(3,1,1);
[count_x,centers_x] = hist(x_abs);
bar(centers_x,count_x);
set(get(gca,'child'),'FaceColor',[0 0.5 0.5],'EdgeColor',[0 0.5 0.5]);
title('X-axes');

subplot(3,1,2);
[count_y,centers_y] = hist(y_abs);
bar(centers_y,count_y);
set(get(gca,'child'),'FaceColor',[0 0.5 0.5],'EdgeColor',[0 0.5 0.5]);
title('Y-axes');

subplot(3,1,3);
[count_z,centers_z] = hist(z_abs);
bar(centers_z,count_z);
set(get(gca,'child'),'FaceColor',[0 0.5 0.5],'EdgeColor',[0 0.5 0.5]);
title('Z-axes');

fprintf('\nProgram paused. Press enter to continue.\n');
pause;
```

## 5.2.2 Representation of the Horizontal Acceleration Magnitude

In the second study the horizontal acceleration magnitude is represented over the time, remember that our ASL algorithm acquires sensor data during 20 seconds every 2 minutes, then the time axis comprehends values between 0 and 20. With this graphic result the difference between the activities is clearer; the wave´s shape illustrates distinct behaviours for each transport.

The features found to represent this behaviour are: Peak Area and Peak Interval Length. The algorithm code will be explained as in the previous section.

Once the data has been opened and assigned to variables the horizontal acceleration value is plotted. Mean value is subtracted from the data array to represent the magnitude around the zero value. Doing the procedure in this way acceleration changes appeared more clearly in the graph, and it is easy to make an analysis of its peaks.

```
%%%% Find Peaks %%%%

%Plot the horizontal acceleration around zero
%Horizontal component depends on how the mobile is oriented,
%so it changes depending on the situation

h_accel = x_acceleration - mean(x_acceleration);
%h_accel = y_acceleration - mean(y_acceleration);
%h_accel = y_acceleration - mean(y_acceleration);

subplot(1,1,1)
accel_plot = plot(timestamp,h_accel,'k-');
hold on;
title('Horizontal Acceleration Graph');
xlabel('time');
ylabel('m/s2');

%define x axis label
datetick('x','ss:fff','keepticks');
```

The next step is finding the main peaks and calculating its area. To achieve the area the *Infinitesimal Integration Method* is used. The process is as follows:

- Find the maximum value [34] in the accelerometer data array, which corresponds to the maximum peak, and mark it in the graph.
- Use a window of 20 samples centered in the peak value.
- Apply the Infinitesimal Integration Method iteratively using the current sample and the subsequent one until the window arrives to its end.
- Repeat the prior instructions 9 times and then the main 10 peaks will be found.

```
%Max values
h_accel_copy = h_accel;
last_peaks = zeros*[1,1];
peak_area = zeros * [1:10];
...

for k = 1:10
    [maximum,pos] = max(h_accel_copy);

    if(abs(pos-last_peaks(:,2))>0)

    last_peaks(k,:) = [maximum,pos];
    max_mark = timestamp(pos);
    mark_max = plot(max_mark,maximum,'rv','MarkerFaceColor','r');
    h_accel_copy(pos) = 0;

    % Finding beginning and ending location

    % Window

    init_peak = pos - 10;
    init_window = timestamp(pos - 10);
    end_peak = pos + 10;
    end_window = timestamp(pos + 10);
```

```matlab
    % Mark the initial and final window points
    yL = get(gca,'YLim');
    hold on;
    init_point = plot([init_window init_window],yL,'r-');
    end_point = plot([end_window end_window],yL,'r-');

    fprintf('\nThe peak begins at %s\n',time(pos - 10));

    fprintf('\nThe peak ends at %s\n',time(pos + 10));

    N = end_peak - init_peak;

    for i=1:N

    peak_area(k) = peak_area(k) + abs(h_accel(init_peak+i-1))*(timing(init_peak+i)-timing(
    init_peak+i-1));

    end
    fprintf('\nThe peak area is %f\n',peak_area(k));

    pause;

    delete(init_point);
    delete(end_point);
    delete(mark_max);

    end
end
```

Finally and average of the peak area is calculated to retrieve as part of the result.

```matlab
average_peak_area = sum(peak_area(:))/10;
fprintf('\nThe average peak are is %f\n',average_peak_area);
```

### 5.2.3  Results for each modality

The numeric and graphic results are shown in the next sections. There is a specific division for the results of each mean.

The units used in the following figures are:

- Acceleration Statistics Results: m/s$^2$
- Peak Interval: seconds (abbreviated as s)
- Peak Area: m/s, since the acceleration integral is the velocity whose unit is m/s

#### 5.2.3.1  Bus

The Bus Analysis has been done over data collected during 13 journeys which are equivalent to 9 hours and 45 minutes travelling by bus.

**Acceleration Components Histogram**

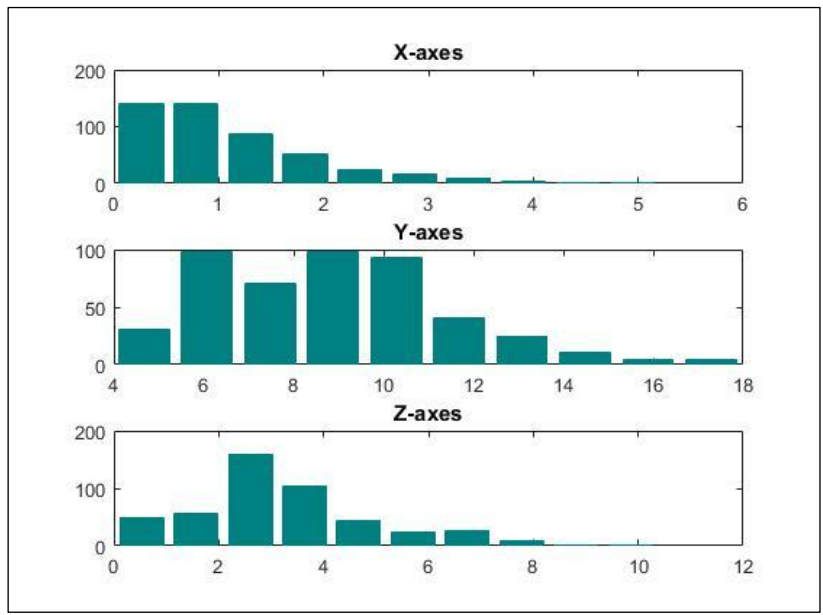

Fig. 5-1: Bus Histogram 1

Histograms show how many sample of each acceleration value has the data analyzed; axis of abscissas represents the acceleration magnitude and axis of ordinates the number of samples for each value.

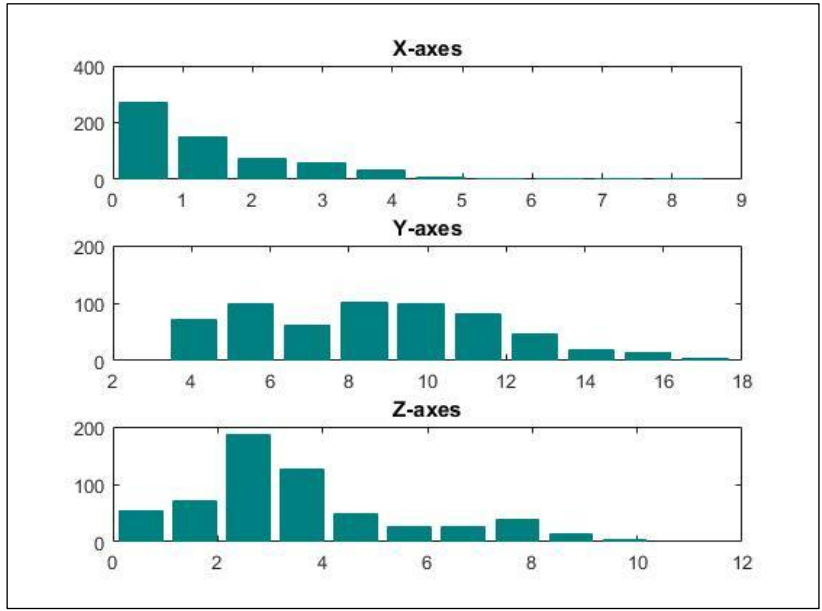

**Fig. 5-2: Bus Histogram 2**

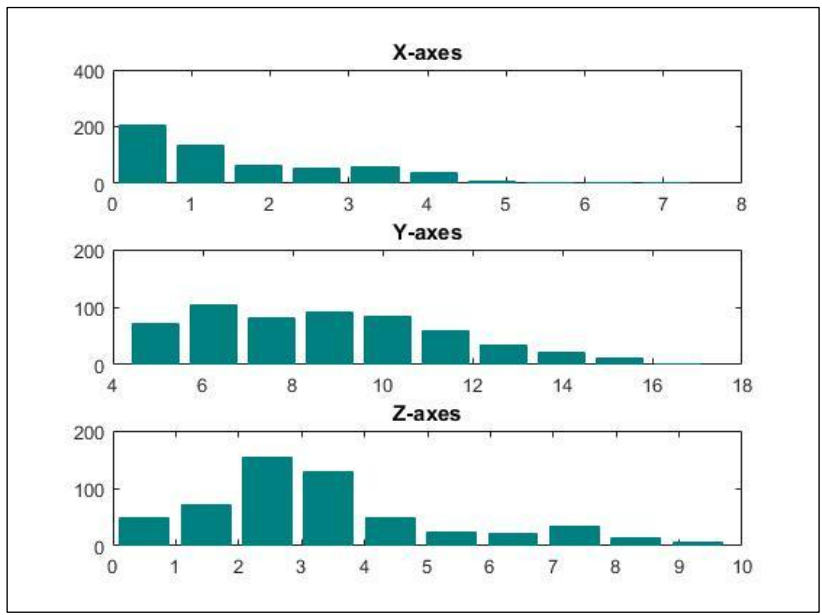

**Fig. 5-3: Bus Histogram 3**

In these samples the gravity is the y-axis components.

The results table shows the average of the results obtained for each parameter.

| | |
|---|---|
| **Mean** | 5,17 m/s² |
| **Standard Deviation** | 0,72 m/s² |
| **Maximum Value** | 8,2 m/s² |
| **Minimum Value** | 0,015 m/s² |

**Table 5-1: Bus Statistics**

## Horizontal Acceleration Representation

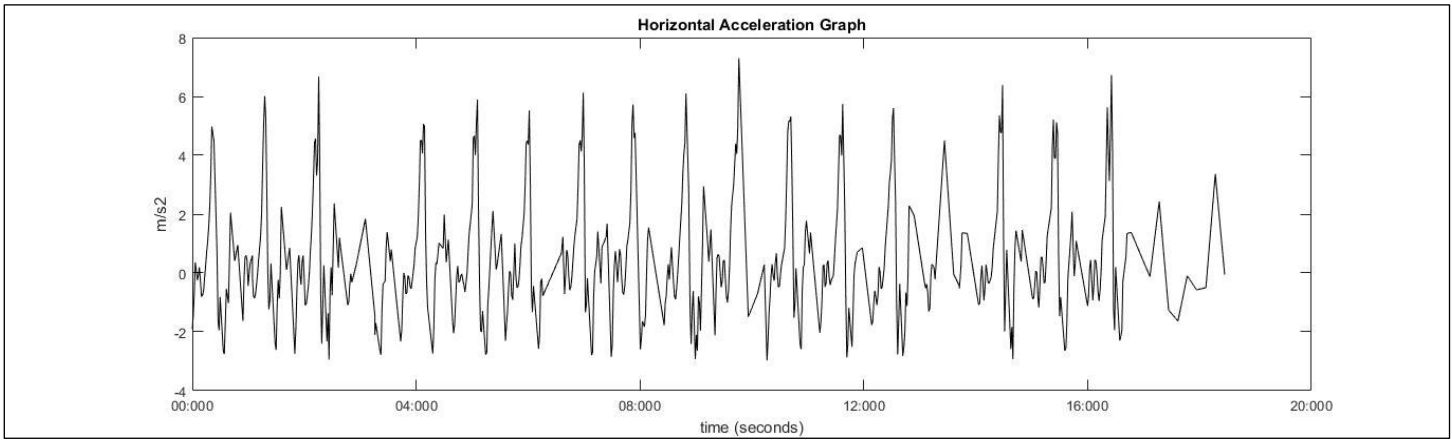

**Fig. 5-4: Bus Horizontal Acceleration 1**

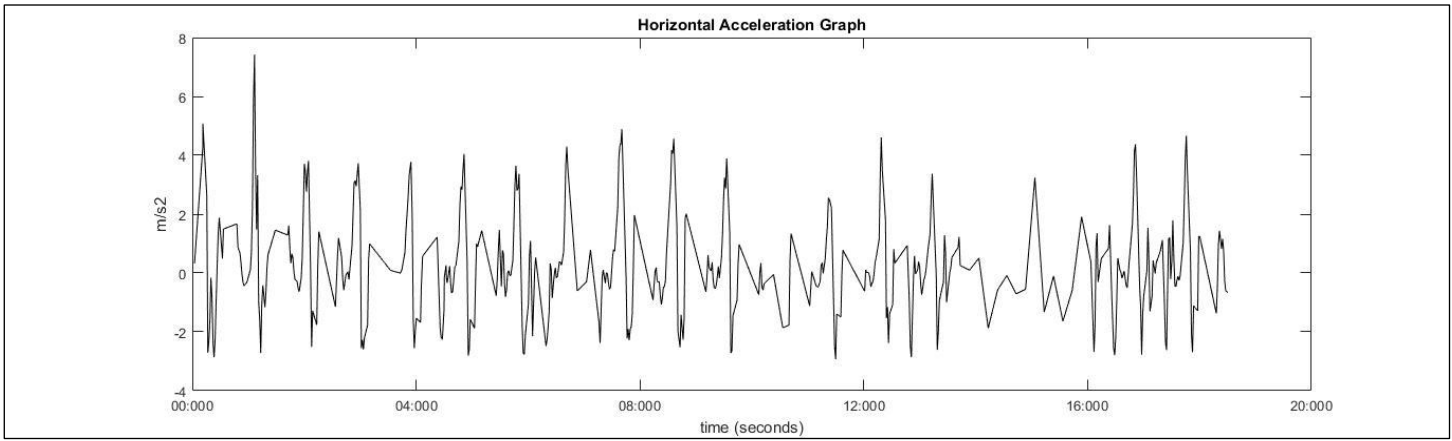

**Fig. 5-5: Bus Horizontal Acceleration 2**

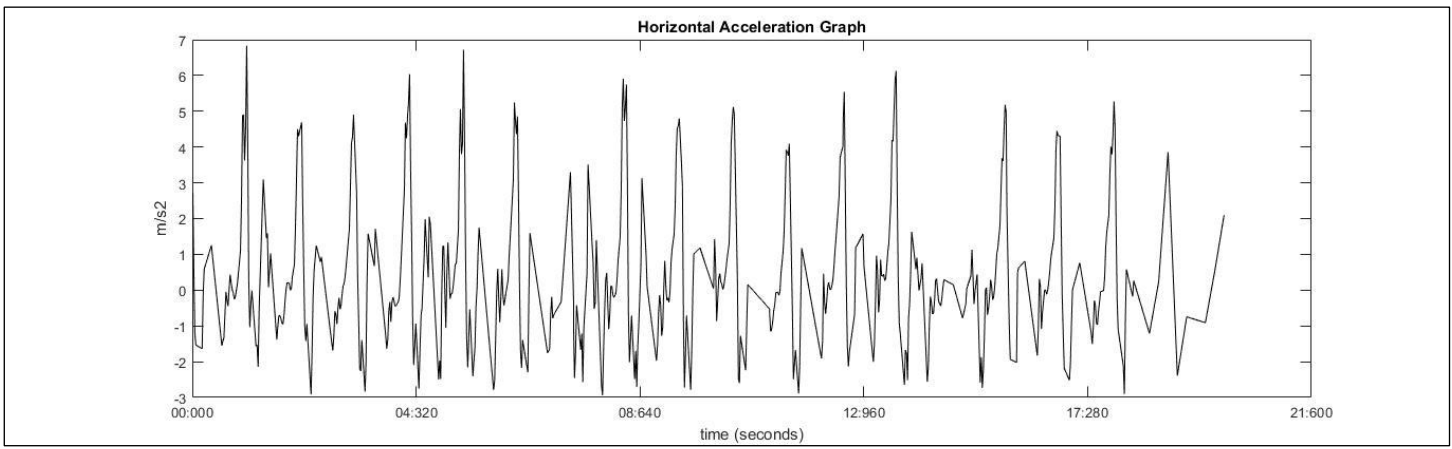

**Fig. 5-6: Bus Horizontal Acceleration 3**

The results table shows the average of the results obtained for each parameter.

| Interval Duration | 0,5 s |
|---|---|
| Peak Area | 1,1 m/s |

**Table 5-2: Bus Peak Parameters**

The Car Analysis has been done over data collected during 10 journeys which are equivalent to 4 hours and 30 minutes travelling by car.

**Acceleration Components Histogram**

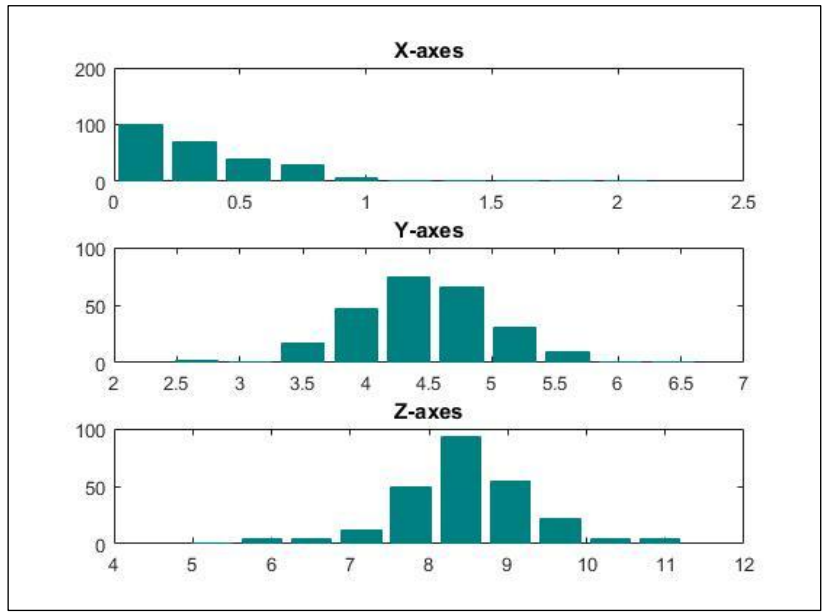

Fig. 5-7: Car Histogram 1

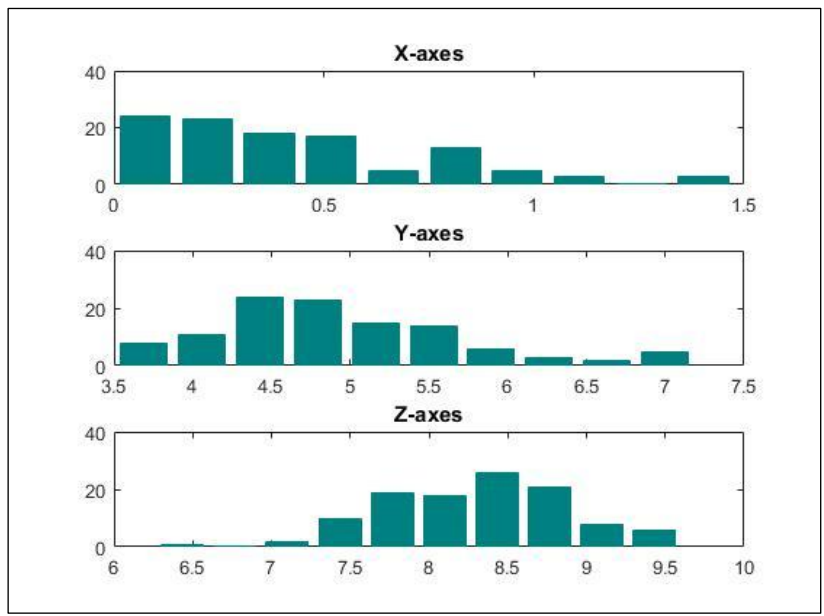

Fig. 5-8: Car Histogram 2

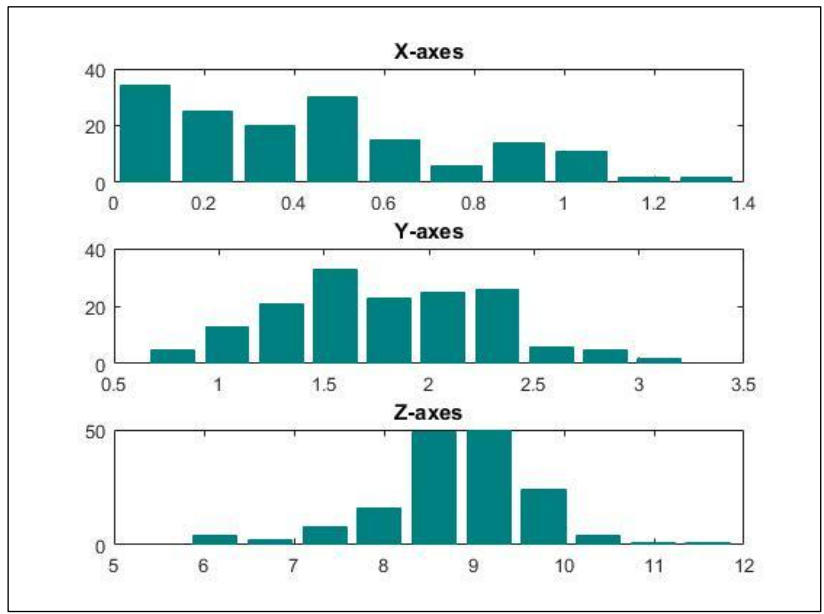

**Fig. 5-9: Car Histogram 3**

In these samples the gravity is the z-axis component.

The results table shows the average of the results obtained for each parameter.

| | |
|---|---|
| **Mean** | **5,26 m/s$^2$** |
| **Standard Deviation** | **0,63 m/s$^2$** |
| **Maximum Value** | **6,14 m/s$^2$** |
| **Minimum Value** | **2,45 m/s$^2$** |

**Table 5-3: Car Statistics**

## Horizontal Acceleration Representation

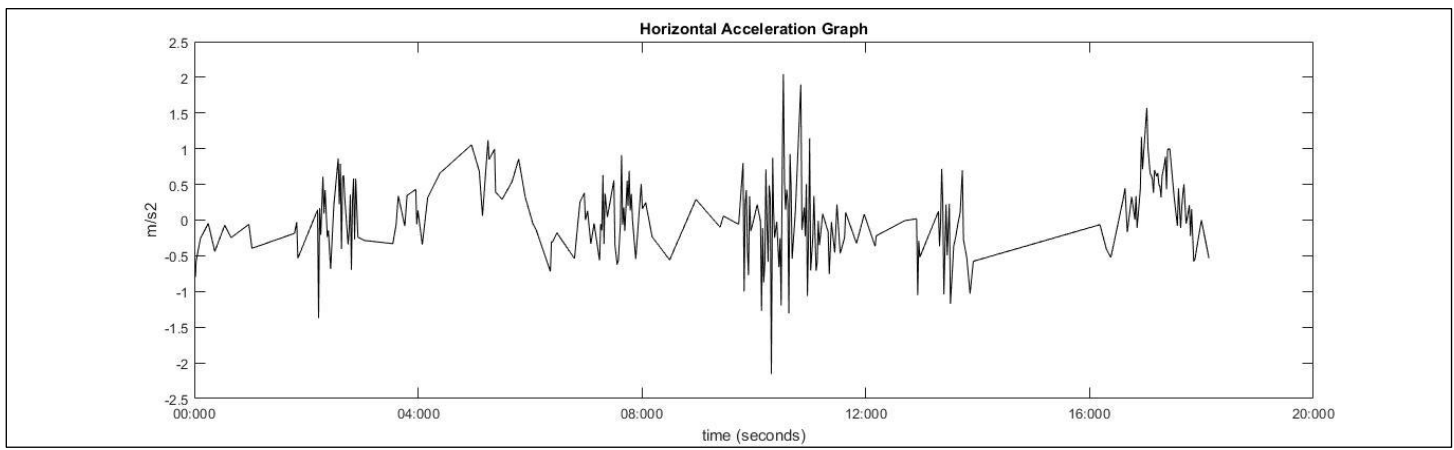

**Fig. 5-10: Car Horizontal Acceleration 1**

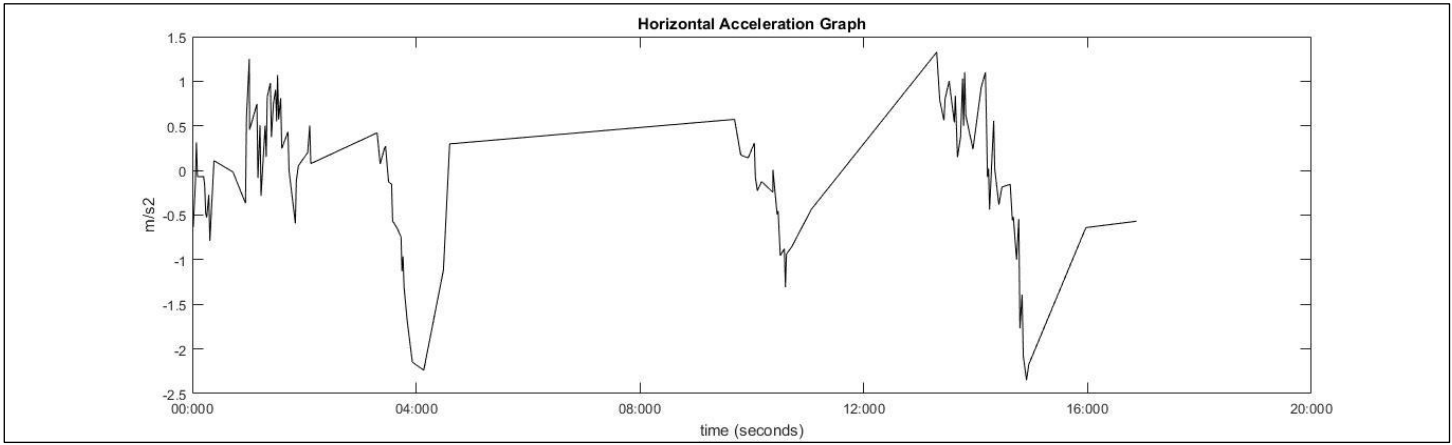

**Fig. 5-11: Car Horizontal Acceleration 2**

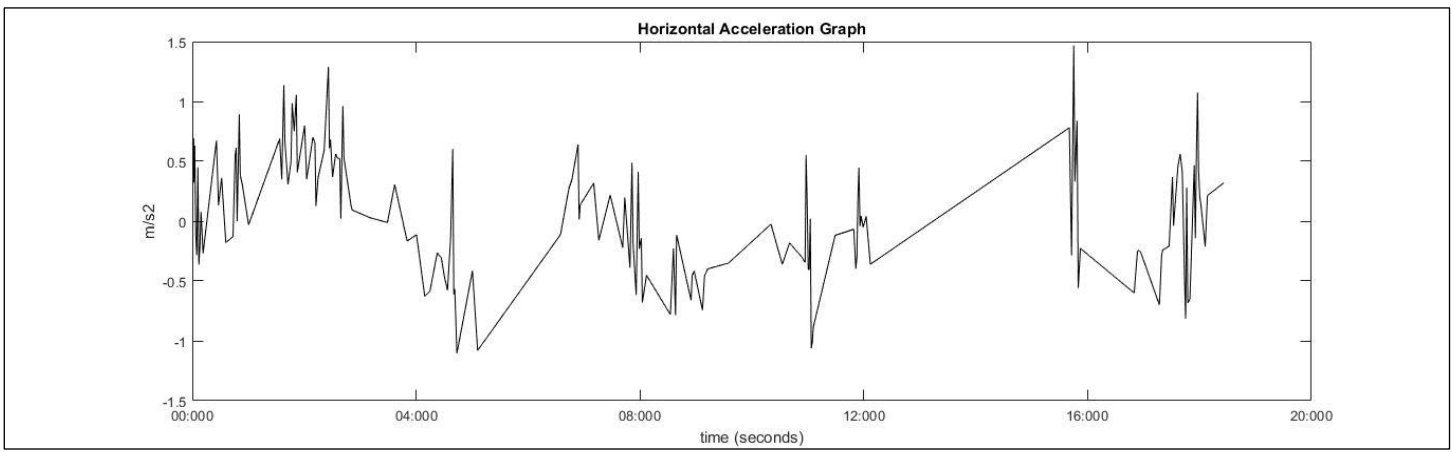

**Fig. 5-12: Car Horizontal Acceleration 3**

The results table shows the average of the results obtained for each parameter.

| Interval Duration | 1,38 s |
|---|---|
| Peak Area | 0,78 m/s |

**Table 5-4: Car Peak Parameters**

### 5.2.3.3    Motorbike

The Motorbike Analysis has been done over data collected during 10 journeys which are equivalent to 4 hours and 40 minutes travelling by motorbike.

**Acceleration Components Histogram**

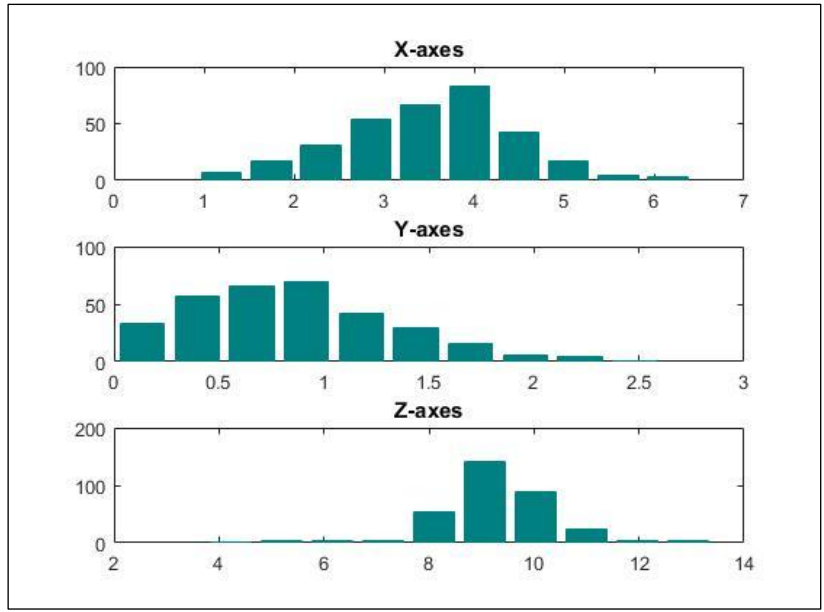

Fig. 5-13: Motorbike Histogram 1

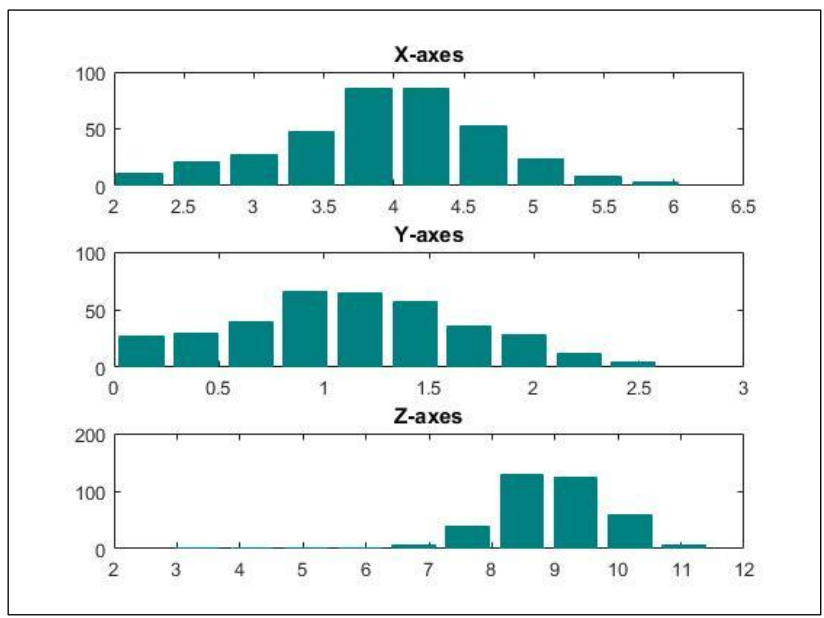

Fig. 5-14: Motorbike Histogram 2

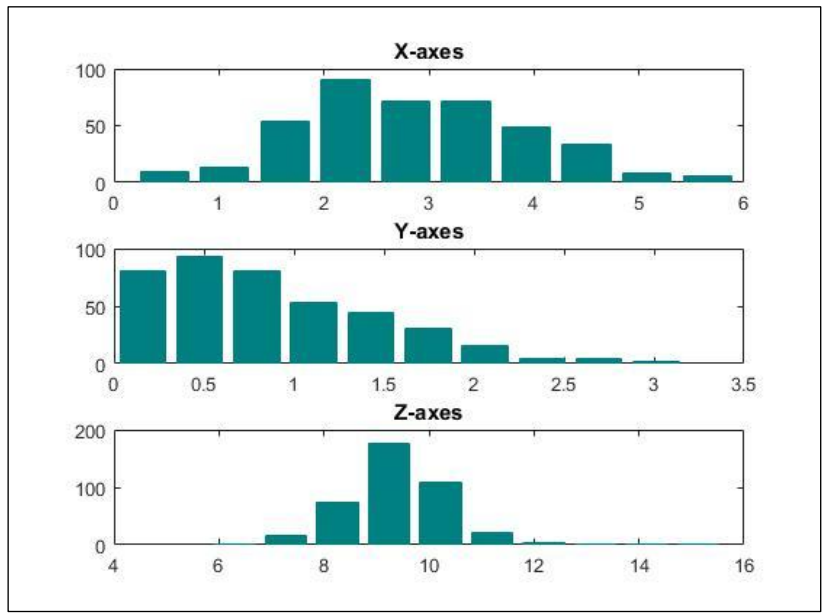

Fig. 5-15: Motorbike Histogram 3

In these samples the gravity is the z-axis component.

The results table shows the average of the results obtained for each parameter.

| Mean | 3,08 m/s$^2$ |
|---|---|
| Standard Deviation | 0,7 m/s$^2$ |
| Maximum Value | 5,55 m/s$^2$ |
| Minimum Value | 1,27 m/s$^2$ |

Table 5-5: Motorbike Statistics

## Horizontal Acceleration Representation

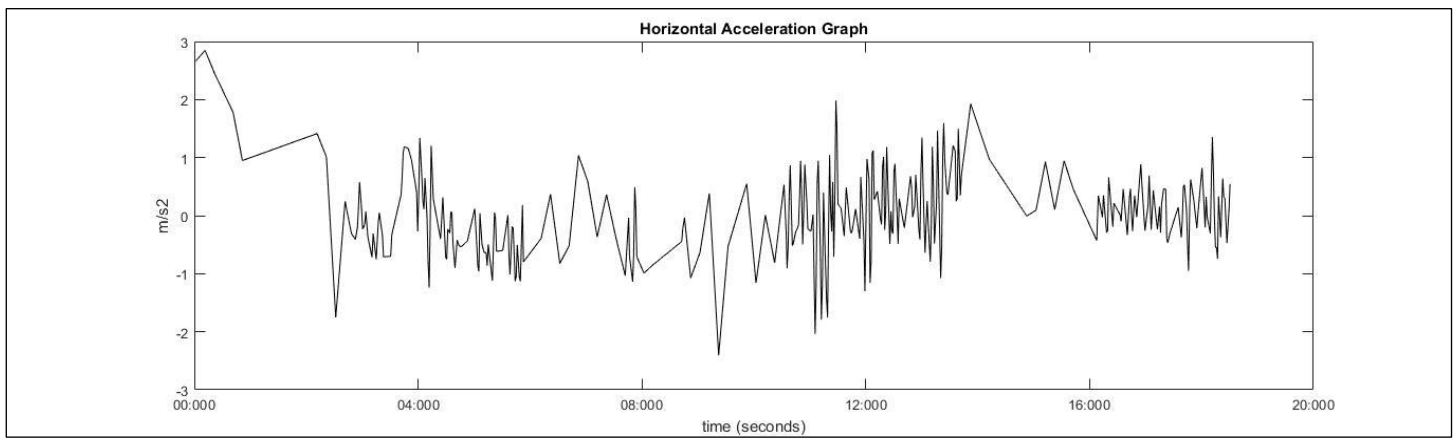

Fig. 5-16: Motorbike Horizontal Acceleration 1

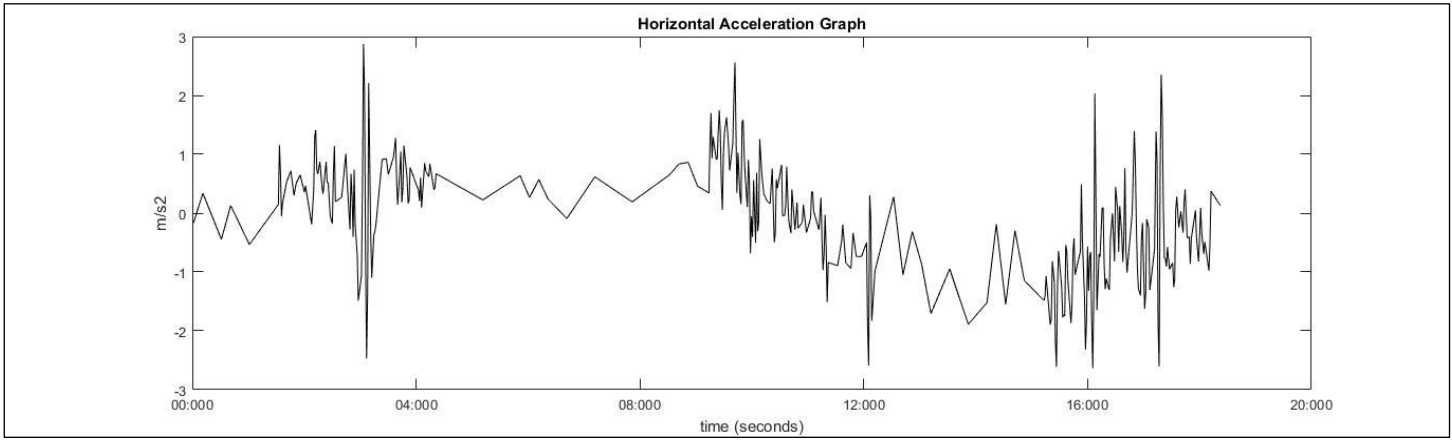

Fig. 5-17: Motorbike Horizontal Acceleration 2

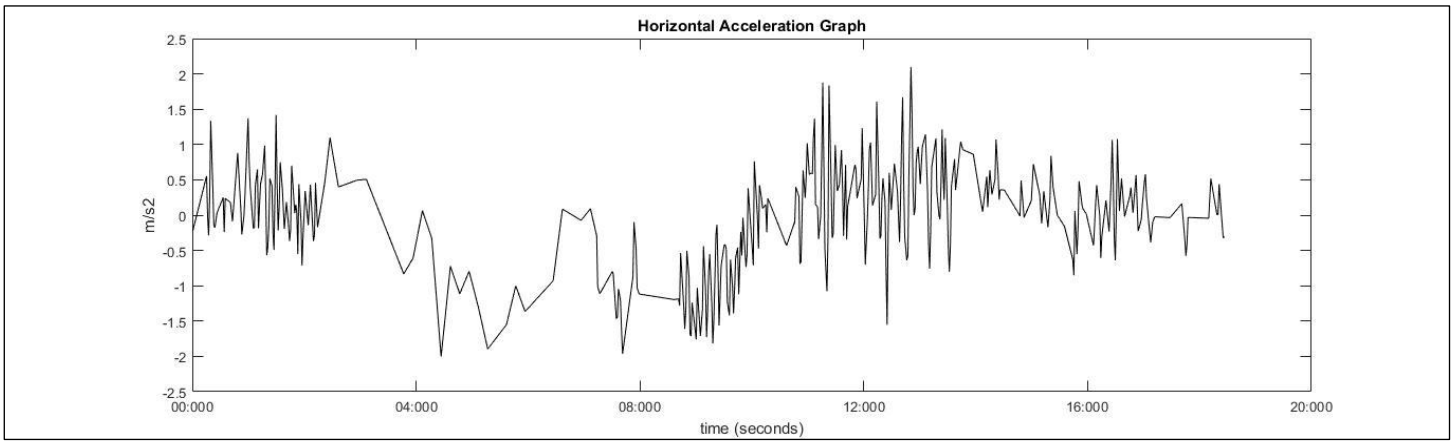

Fig. 5-18: Motorbike Horizontal Acceleration 3

The results table shows the average of the results obtained for each parameter.

| Interval Duration | 0,52 s |
|---|---|
| Peak Area | 0,66 m/s |

Table 5-6: Motorbike Peak Parameters

### 5.2.3.4    Metro

The Metro Analysis has been done over data collected during 17 journeys which are equivalent to 5 hours and 40 minutes travelling by metro.

#### Acceleration Components Histogram

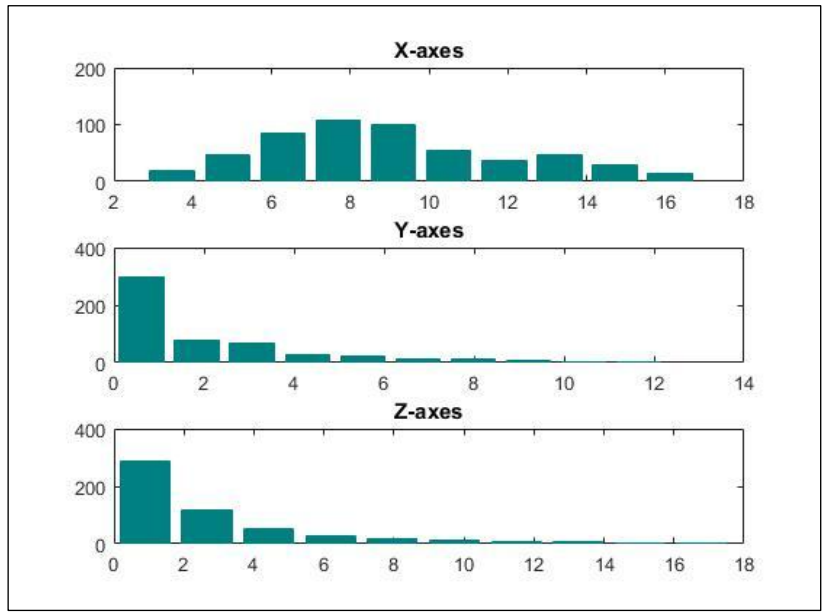

Fig. 5-19: Metro Histogram 1

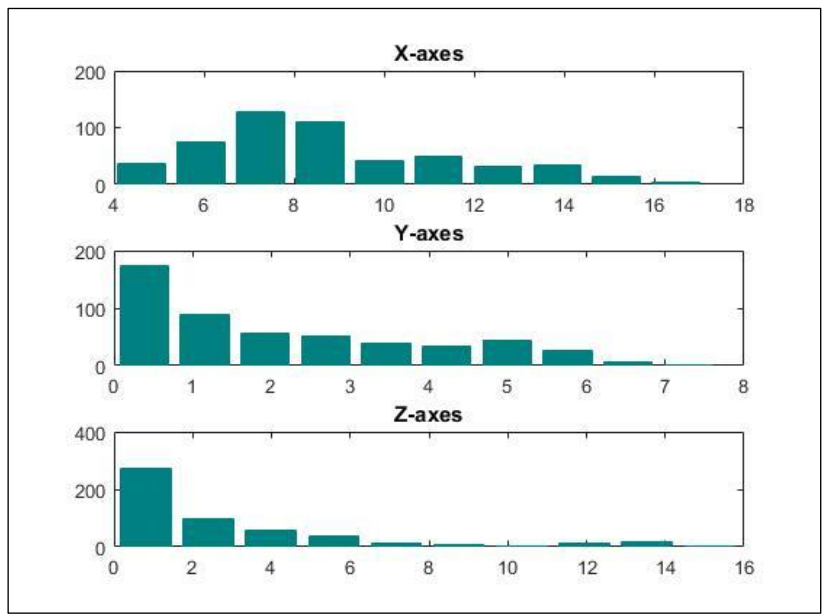

Fig. 5-20: Metro Histogram 2

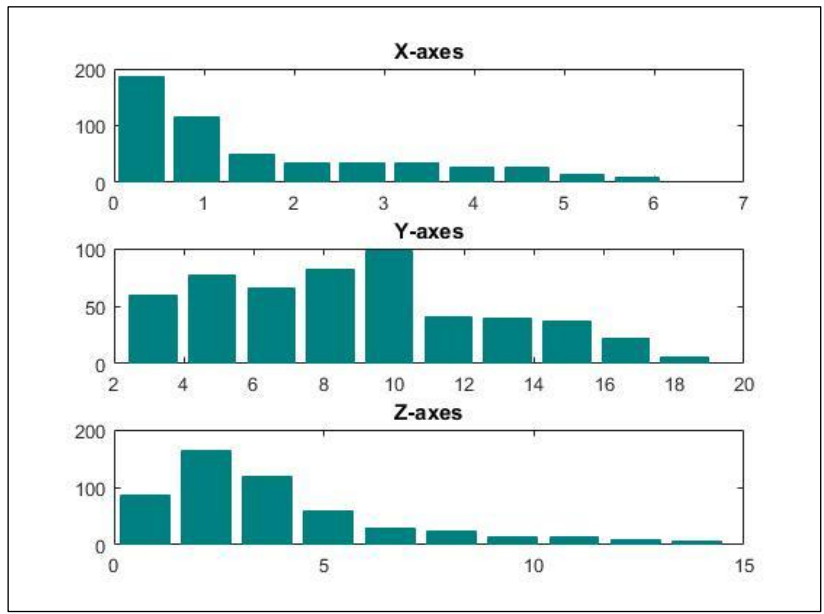

Fig. 5-21: Metro Histogram 3

In the first two samples the gravity component is the x-axis, but the last one it is the z-axis.

The results table shows the average of the results obtained for each parameter.

| Mean | 3,32 m/s$^2$ |
|---|---|
| Standard Deviation | 2,52 m/s$^2$ |
| Maximum Value | 13,7 m/s$^2$ |
| Minimum Value | 0,04 m/s$^2$ |

Table 5-7: Metro Statistics

## Horizontal Acceleration Representation

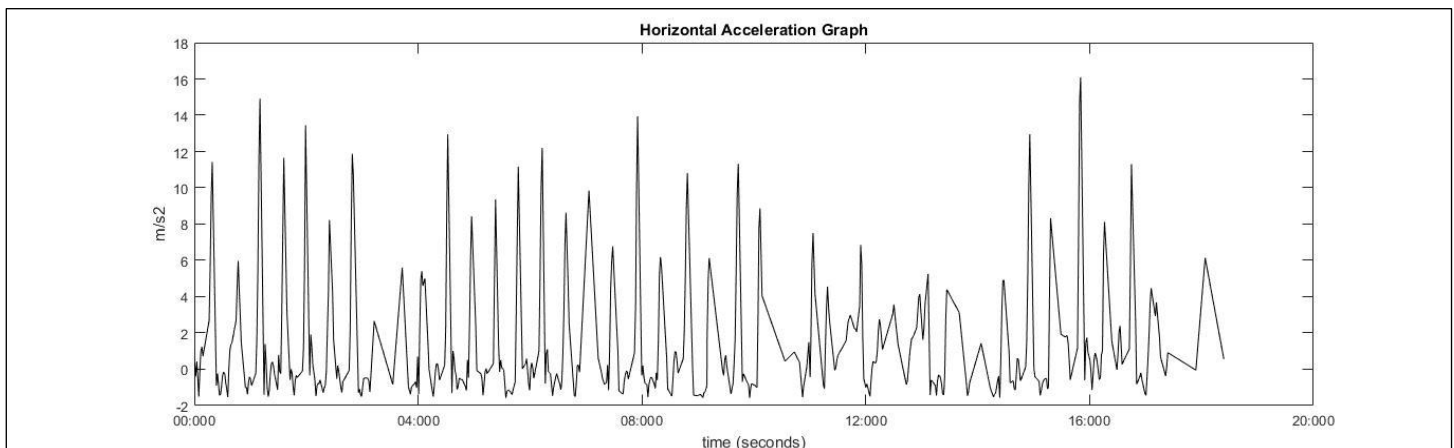

Fig. 5-22: Metro Horizontal Acceleration 1

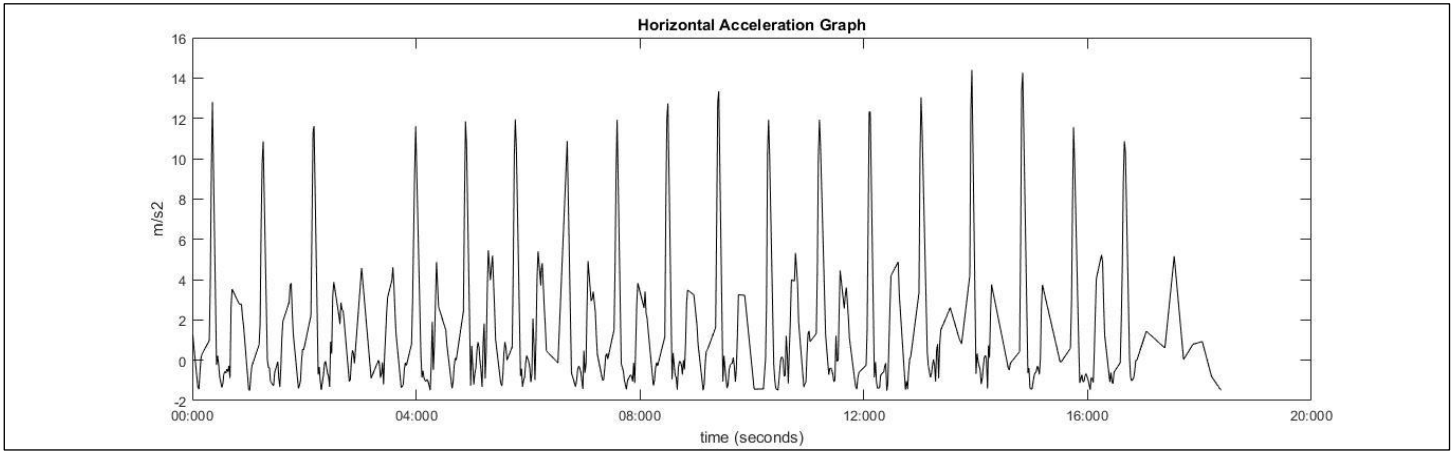
**Fig. 5-23: Metro Horizontal Acceleration 2**

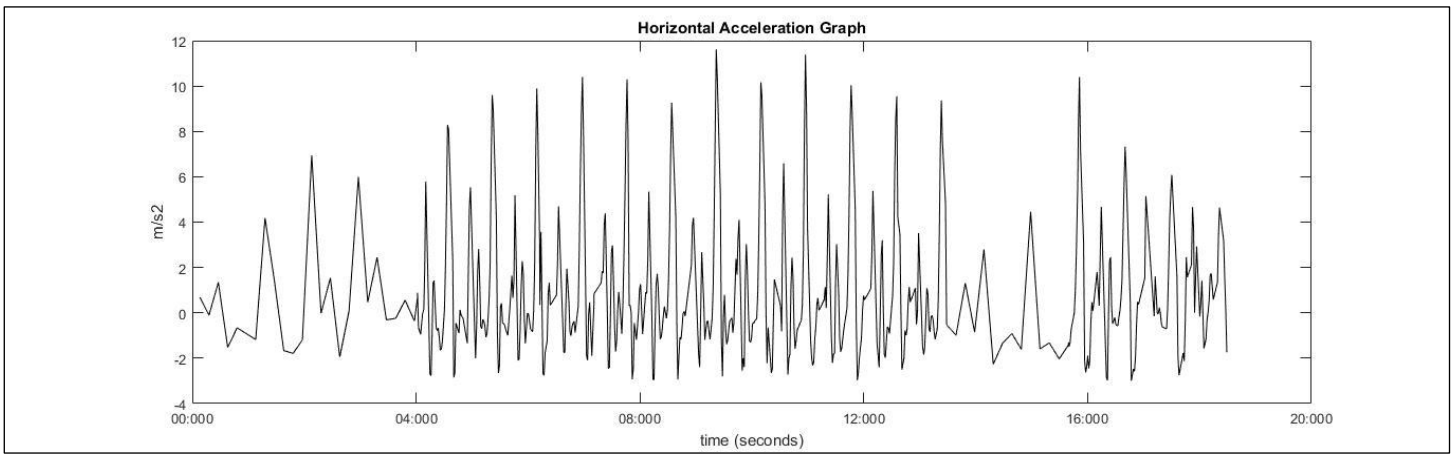
**Fig. 5-24: Metro Horizontal Acceleration 3**

The results table shows the average of the results obtained for each parameter.

| Interval Duration | 0,6 s |
|---|---|
| Peak Area | 1,57 m/s |

**Table 5-8: Metro Peak Parameters**

### 5.3.2.5    Train

The Train Analysis has been done over data collected on a round trip between Barcelona and Cartagena which are equivalent to 5 hours and 40 minutes travelling by train.

**Acceleration Components Histogram**

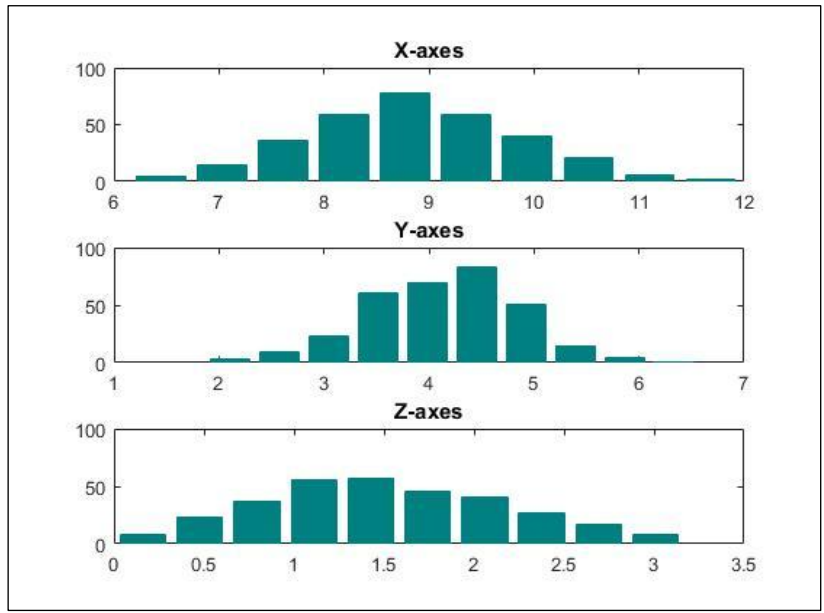

Fig. 5-25: Train Histogram 1

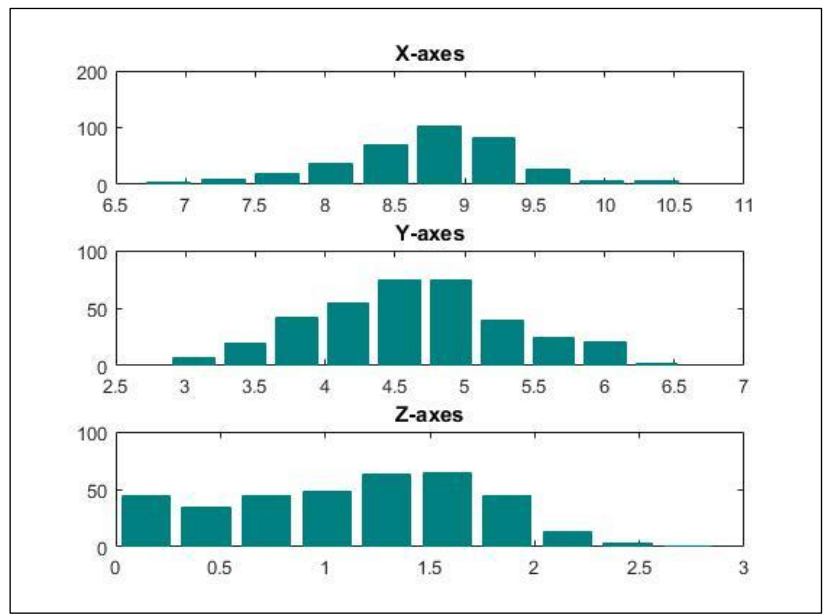

Fig. 5-26: Train Histogram 2

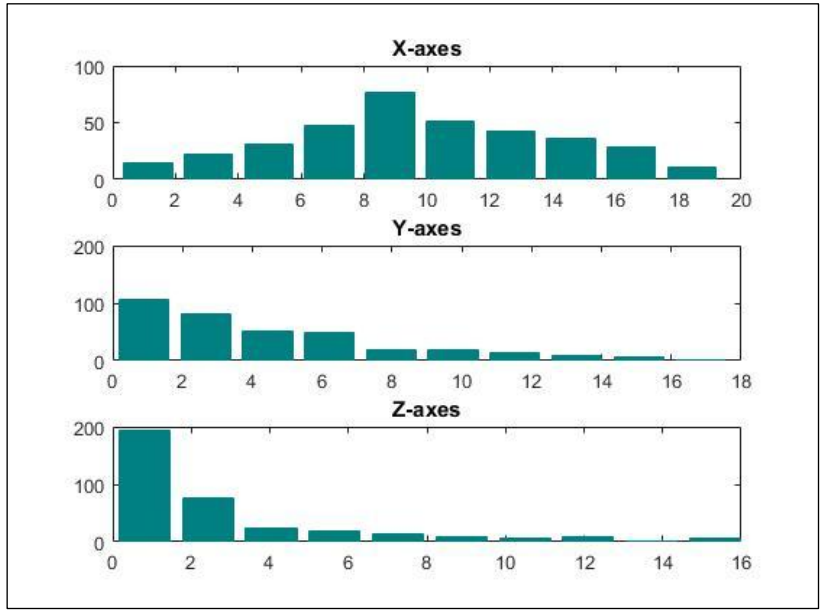

Fig. 5-27: Train Histogram 3

In these samples the gravity is the x-axis component.

The results table shows the average of the results obtained for each parameter.

| Mean | 4,13 m/s$^2$ |
|---|---|
| Standard Deviation | 2,41 m/s$^2$ |
| Maximum Value | 12,36 m/s$^2$ |
| Minimum Value | 0,09 m/s$^2$ |

Table 5-9: Train Statistics

## Horizontal Acceleration Representation

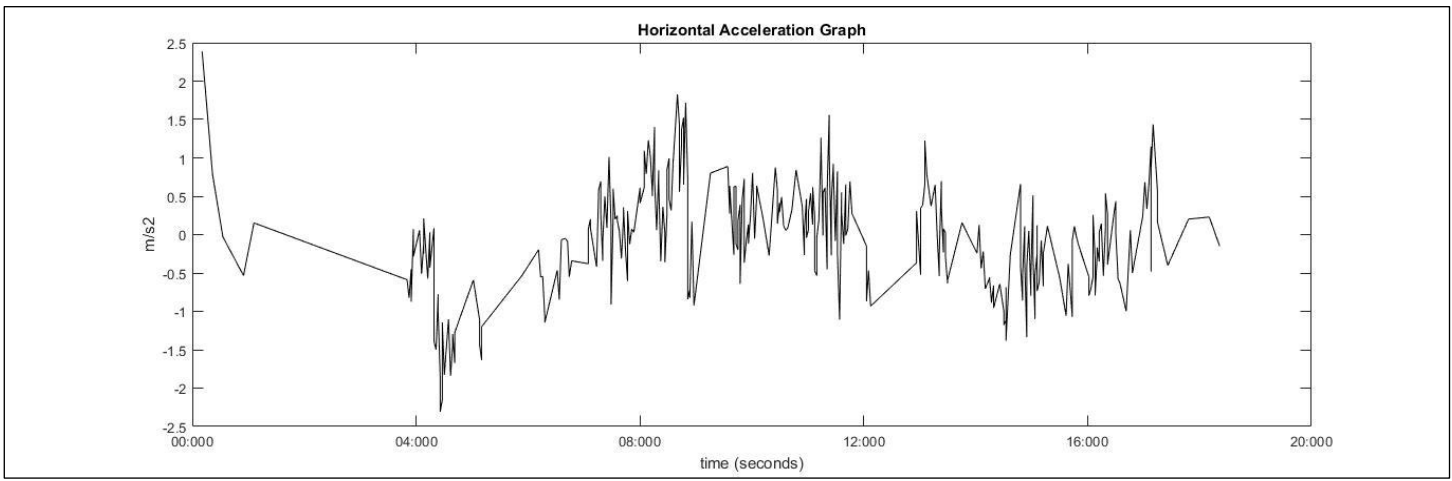

Fig. 5-28: Train Horizontal Acceleration 1

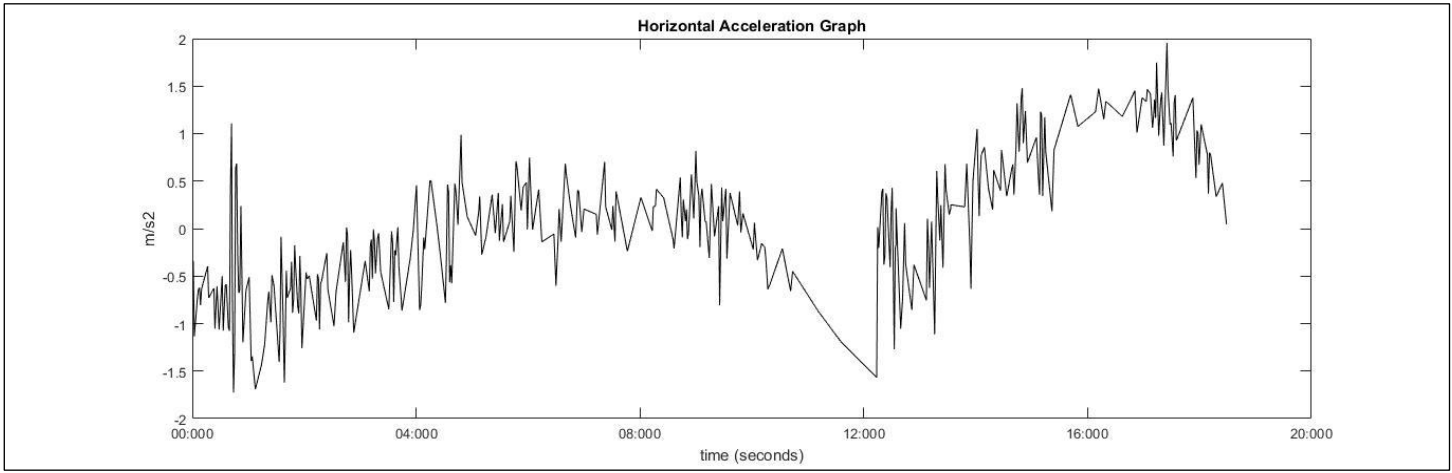

**Fig. 5-29: Train Horizontal Acceleration 2**

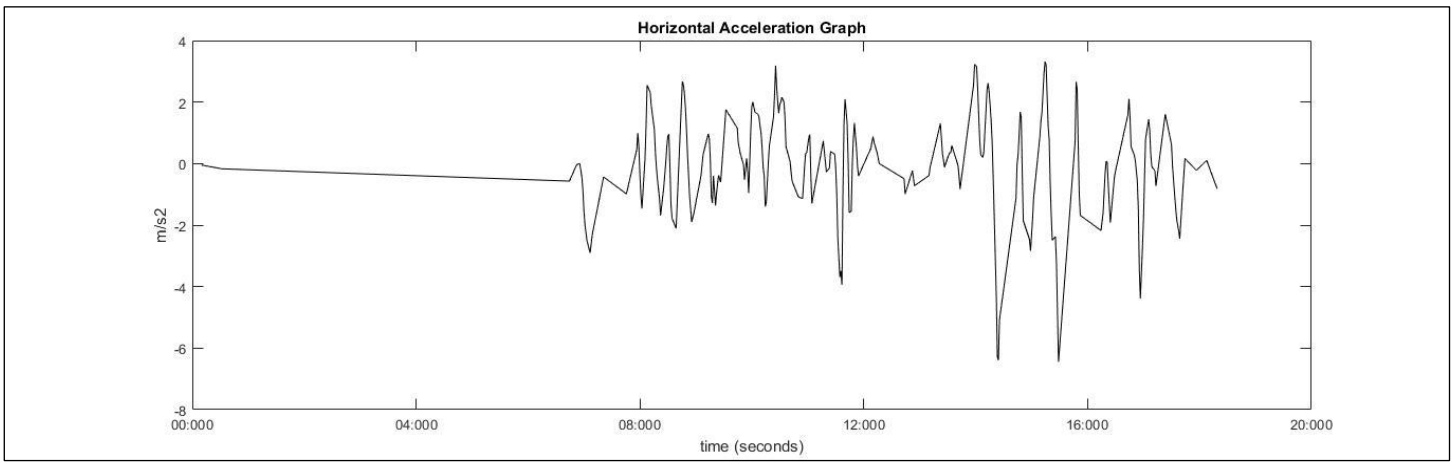

**Fig. 5-30: Train Horizontal Acceleration 3**

The results table shows the average of the results obtained for each parameter.

| Interval Duration | 0,67 s |
|---|---|
| Peak Area | 1,65 m/s |

**Table 5-10: Train Peak Parameters**

## 5.3    Mobility Patterns

### 5.3.1  Interpretation of the results

Looking at the previous Horizontal Acceleration Graph figures, it is clear that each modality has its own wave shape. This supports the initial idea suggested by the developer team, which was to find patterns to model each activity and use it to detect the transportation mode. The next tables are a compilation of the statistical and peak features found in the previous section with an accurate interpretation of the results.

|  | Mean | Standard Deviation | Maximum | Minimum |
|---|---|---|---|---|
| **Bus** | 5,17 | 0,72 | 8,2 | 0,015 |
| **Car** | 5,26 | 0,63 | 6,14 | 2,45 |
| **Motorbike** | 3,03 | 0,7 | 5,55 | 1,26 |
| **Metro** | 3,32 | 2,52 | 13,7 | 0,004 |
| **Train** | 4,13 | 2,41 | 12,36 | 0,097 |

**Table 5-11: Statistics Parameters**

As it can be seen in the first table the statistic parameter Standard Deviation classifies the vehicles in two modalities; those which operate on roads and those operating on rails. A first classifier would be implemented based on this property.

|  | Interval Length | Peak Area |
|---|---|---|
| **Bus** | 0,5 | 1,1 |
| **Car** | 1,37 | 0,78 |
| **Motorbike** | 0,52 | 0,66 |

**Table 5-12: Road Peak Parameters**

|  | Interval Length | Peak Area |
|---|---|---|
| **Metro** | 0,6 | 1,57 |
| **Train** | 0,67 | 1,65 |

**Table 5-13: Rail Peak Parameters**

The second table shows, not as clear as the first one, differences inside the vehicle´s engine modality classification.  On the road vehicles side Bus is the one that has the large peak area and this could be used to distinguish it from the other two. On the other hand, Car and Motorbike would be distinguished using the interval length, which is longer on Car and less than the half of this value on Motorbike. Rail Peak Features do not seem to distinguish between its modalities; at least it does not under developers' criterion.  For this reason in future works the difference between metro and train will be studied accurately.

### 5.3.2  Classification Diagram

The following diagram shows how would be the workflow a possible solution that would implement the patterns found in the current study.

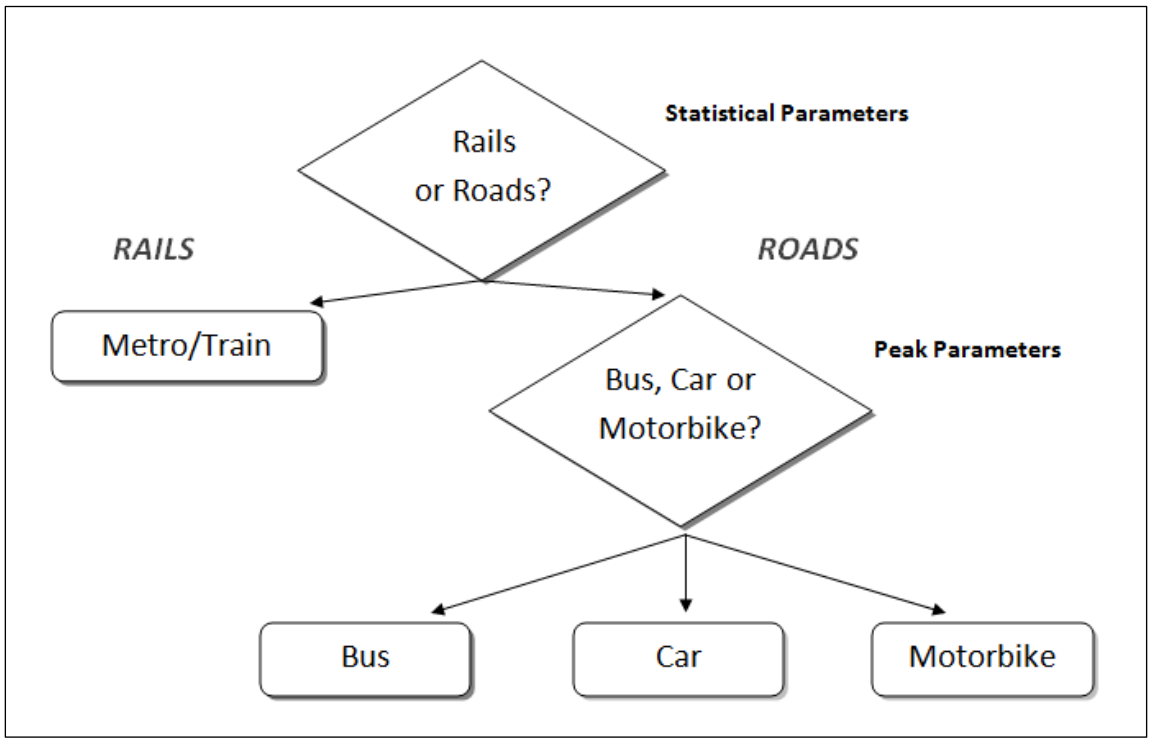

Fig. 5-31: Classification Diagram 1

Peak parameters classifier could be implemented in the way illustrated below.

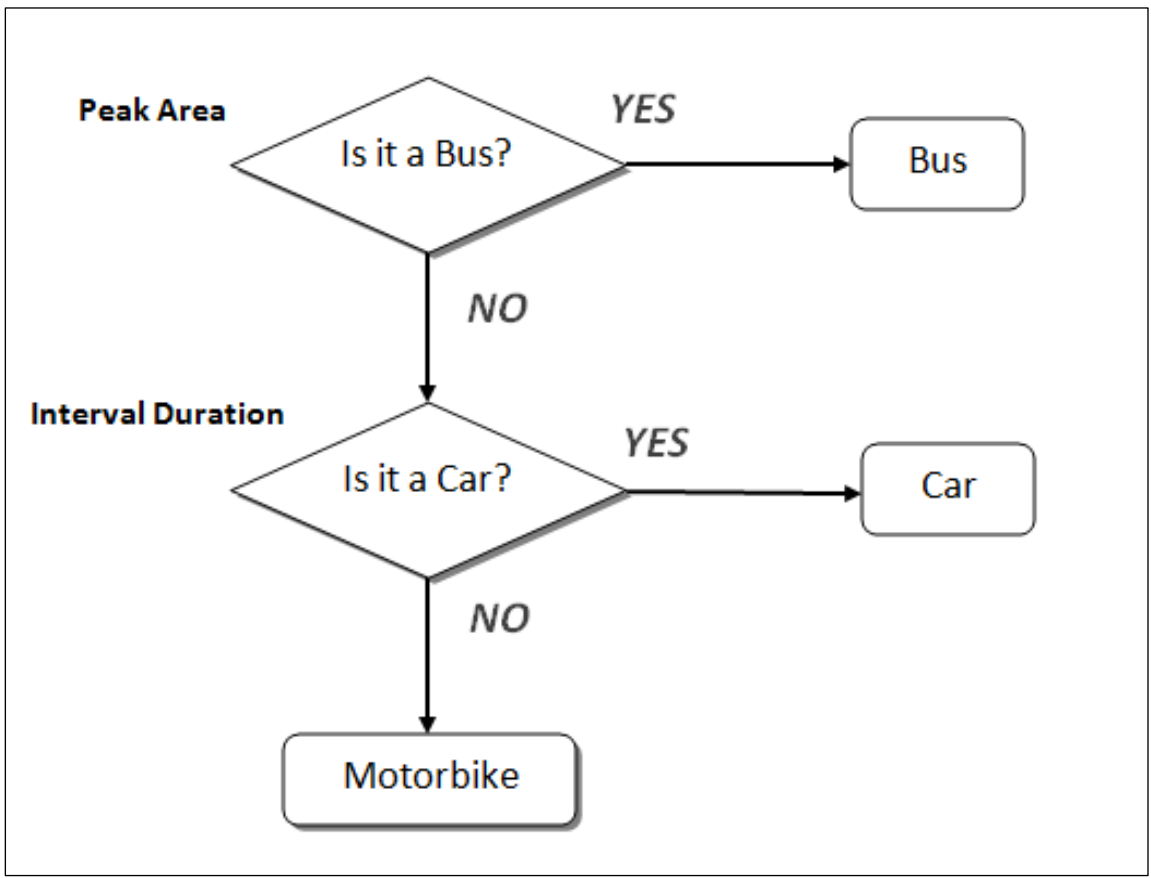

Fig. 5-32: Classification Diagram 2

# 6.  Conclusions & Future Work

## 6.1     Conclusions

The developed algorithm allows listening the accelerometer and gyroscope sensors which can provide an amount of data related to the user´s activity. Accelerometer Sensor Listener is a scalable solution; the information obtained from its use offers many possibilities depending on the final service. In the current research it has been a tool for obtaining mobility data from the different transport modalities.

The analysis after the data collection process has reached its objective; the result is patterns to distinguish the citizen´s activity. Each transport mean has its own behaviour and the current study demonstrates it, then implementing an algorithm that uses these parameters would be a solution for transportation mode detection task instead the use of Google API Services.

Mobility Area, inside the Smart City concept, is the context of the project, so the purpose is providing smart solutions to the issues in a daily journey. Furthermore, mobility information is truly appreciated.

## 6.2     Future Work

MobilitApp is currently in development, adding some new features and implementing awesome functionalities.

- An **E-Call**; taking the advantage of monitoring the accelerometer an emergency call is being developed. The application would detect possible falls or accidents and a notification would be sent to emergency services telephone number (e.g., 112 in Europe or 911 in North America).

- Continuing with the line of this project, the implementation of the **transport mode detection algorithm** using the results of the mobility analysis and **machine learning** techniques would be accomplished. All the data collected during this project could be used as the training set.

- Combine MobilitApp service with an app which would **attract more users**. The proposal is designing an eco-friendly game application that categorizes with a kind of badge the citizens´ behaviour depending if they use public or private transport, bicycle or motorbike, and in this way a high level of interaction would be achieved.

- When the application audience will increase the **project´s infrastructure** should be adapted to the new needs; higher processing power, more memory for database, and a more reliable encryption system.

# 8.  Annexes

## Annex A – Setting a server up with Raspberry Pi

1. **Introduction**
   - 1.1. Hardware and Software
   - 1.2. Server's purpose and aim of the guide.

2. **Raspbian Operating System**
   - 2.1. Installing Raspbian in SD cards.
   - 2.2. Execution and configuration of the Operating System.
   - 2.3. Graphics desktop.

3. **Defining static IP address for our Raspberry.**

4. **Server accessing from the outside.**
   - 4.1. Port Forwarding
   - 4.2. Dynamic DNS
   - 4.3. Safety in the server

5. **LAMP configuration in our server**

6. **Server's use**
   - 6.1. Web
   - 6.2. Data Base

# 1. Introduction

On this guide we will explain how to set the server using a Raspberry Pi. The Raspberry Pi is a small chip computer; the design consists of a microprocessor, a RAM memory and entry points for peripherals, all of them integrated into a single small size chip. Using a Raspberry is an advantage for the server setting if we have limited economic resources.

The explanation is composed of several steps that we will try to expose with some detail, as well as define the protocols and technical aspects that are involved in the process.

## 1.1. Hardware and Software

### Hardware

- *Raspberry Pi*, in this case the used device is *Raspberry Pi 2 Model B*.
- MicroSD card, 64Gb class 4.
    Recommended minimum capacity for the image support of an Operating System is 4Gb. The class indicates the data transfer speed the card can support.
- Ethernet wire.
- USB 1.8 A/5V adapter.

### Software

- Raspbian Operating system image.

## 1.2. Server's purpose and aim of the guide.

We are a group of students from the Polytechnic University of Catalunya who are together developing a project based in a mobile application. The purpose of this server is to provide the services for that application, as well as to store its data base and web page.

The aim of this guide is to help anyone making similar projects to ours, which require of an economical server to develop their ideas or who want to begin using a Raspberry Pi.

# 2. Raspbian Operating System

Raspberry devices use Linux Operating Systems. The one used by us is Raspbian, a distribution from Debian optimized to be used with Raspberry hardware and which we can obtain from the official Raspberry web site.

## 2.1. Installing Raspbian in SD cards.

We download the Raspbian image from the official Raspberry web site:

[https://www.raspberrypi.org/downloads/raspbian/](https://www.raspberrypi.org/downloads/raspbian/)

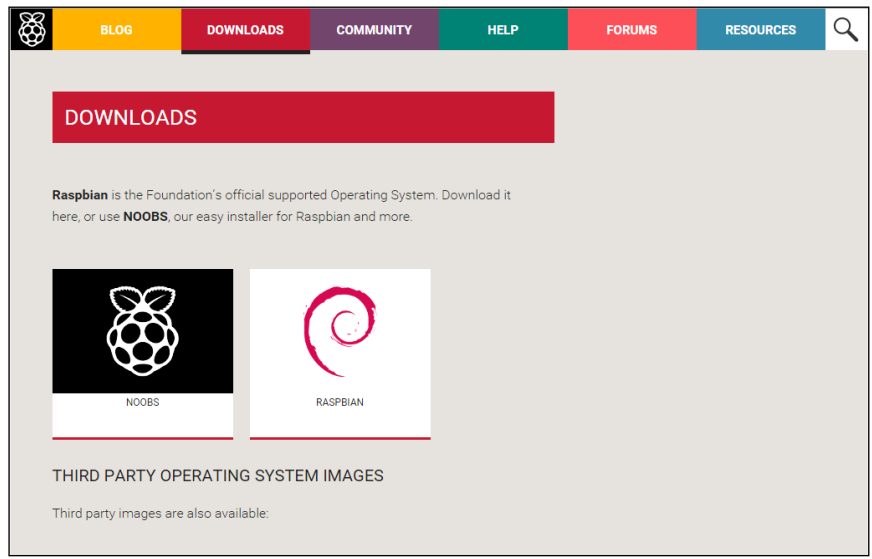

Once finished the download, we install the image in the card, but for that we will use a tool to make that task. Win32DiskImager is a software that allows the installation of operating system images into USB devices and SD cards. We can download it from the following link:

[http://sourceforge.net/projects/win32diskimager/](http://sourceforge.net/projects/win32diskimager/)

We now have all the software needed for the installation of Raspbian in the microSD card.

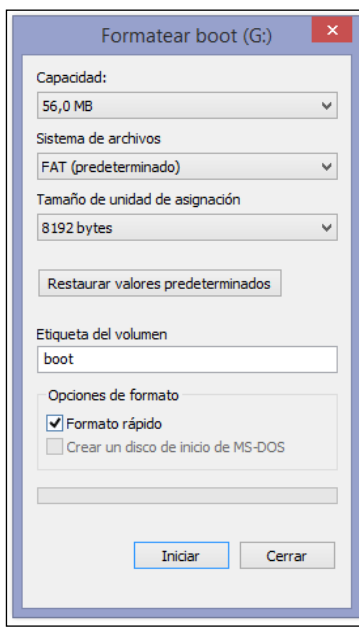

The previous step to the installation is to make sure that the SD card is free of content. If the card has been bought specifically for this purpose, we can skip this step, otherwise, we should format it.

We must then insert the card on our PC reader and once it is recognized we must then open *Devices and Unities in My Computer*. We found then our SD card, click with the right button on it and we select to format.

Now that we have formatted the card, the next step is to install the operative system into it.

We open the Win32DiskImager programme as administrator; select the Raspbian image on the pull-down menu and the microSD card on the device.

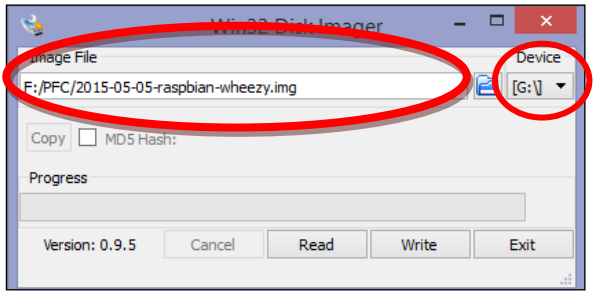

We then press on writing and wait around 2-3 minutes, which is what it usually takes to install.

## 2.2. Execution and configuration of the Operating System

When the installation is finished we take out from thecomputer reader the microSD card and we insert it on the appropriate Raspberry groove. Now we connect the Raspberry to the router through Ethernet and finally we plug it to the power grid (recommended on this order).

To gain acces to the Raspberry we will connect using a connection SSH.[1]

We download the SSH *Putty* client on our computer. There are more SSH clients that allow the distant connection to other devices, the more convenient one can be chosen.

On the following link we can find the *Putty* client:

http://www.chiark.greenend.org.uk/~sgtatham/putty/download.html

When we already have the software that will let us do the distant connection, there is only left to know the IP address of the computer to which we want to connect. For that purpose we must connect to our router, so we open a terminal and check our port of linkage (router).

---

[1] SSH (Secure Shell): this is a protocol for the level of application used to gain access distantly to devices using a net, allowing to control those devices with a command interpreter.

Using the command *ipconfig* we can see the configuration of the wire to which our computer is linked.

Found the IP of our router, we open a search engine and paste this address.

On the configuration section of the net, we will see a table with the IP addresses that are being used. The section where this information is found will depend on the router model but the difference that can be found will be minimum.

We can see on the previous table that there are four addresses assigned, the first three are from other home computers (we can check this by writing the command *ipcofing* in a terminal), so our Raspberry device is the 192.168.1.92 address.

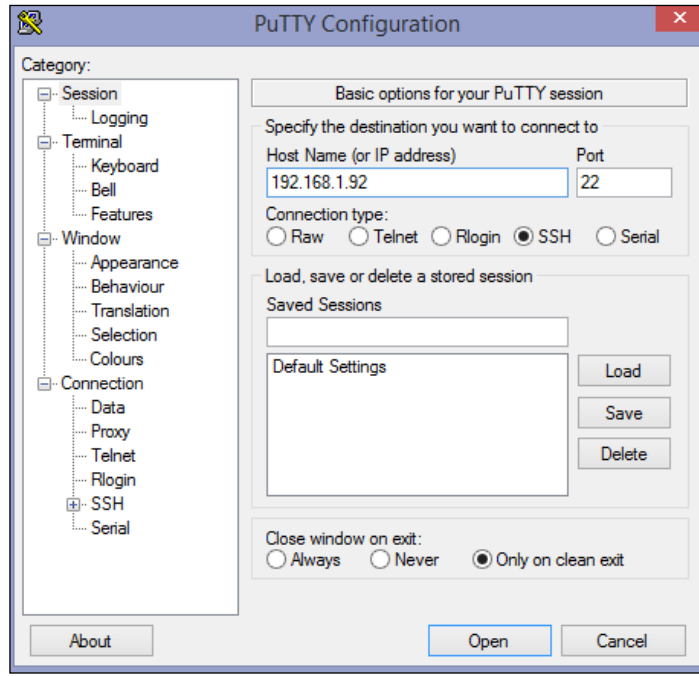

Then the Putty must be opened and we specify what kind of connection will be done (SSH), the IP address or host name (192.168.1.92) and the port (22).

To be able to enter the Raspberry, we are asked for a username and a password, which by default is: *user: pi*, *password: raspberry*.

Once we are inside the Raspberry Pi, we can start making the first configurations.

*sudo raspi-config*

Using the previous command, the raspbian composition tool is opened, which allows us to do a first system configuration.

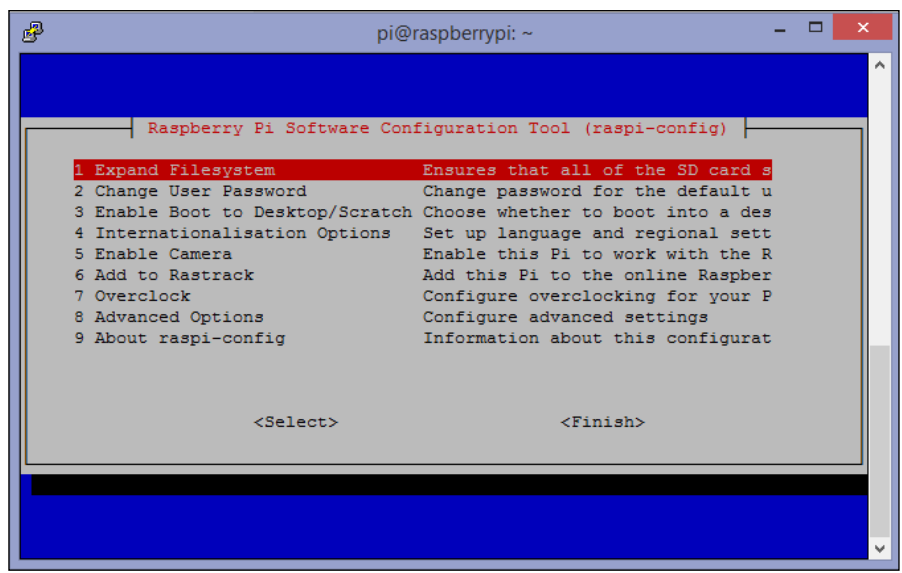

Next to each option, we will find a short explanation of what we will be able to modify. The first ones we are going to adjust will be:

**Expand Filesystem:** so that there is not unused space on the SD we expand the operating system to all the capacity of the card.

**Internationalisation Options:** to change the language, the time zone and the keyboard format.

In order to end with this part, we will update the Firmware and so we will guarantee the appropriate operation of the hardware in our Raspberry.[2]

*sudo rpi-update*

Reset for the update to come into effect.

*sudo reboot*

## 2.3. Graphics desktop

When we are working distantly on a device, we develop all the activities through terminal, we do not have direct access to the desktop. In the event of being interested on gaining access to the desktop of that machine, we should install a graphics desktop server on it.

This step is not essential for the setting of the server, but could be beneficial to explain it in any case.

**VNC (Virtual Network Computing)**

VNC is a graphics desktop that allows controlling the Raspberry Pi desktop remotely from our computer.

We must install the package *Tight VNC* in the Raspberry; it basically consists of the server that will be run in the device each time that achieving a graphical connection to it is desired.

*sudo apt-get install tightvncserver*

When strating the VNC session for the first time, we are asked for a password; this password will be asked every time we connect to the Raspberry from a different client, on this case our own PC.

*vncserver :0 (when starting session in the server we must say which monitor we use: 0,1,2… )*

*vncserver:0  -geometry 1920x1080 - depth 24 (the option of specifying the size of the monitor is also available)*

---

[2] Firmware: this is a low level instruction group that sets the logic that rules the electrical circuits of a device.

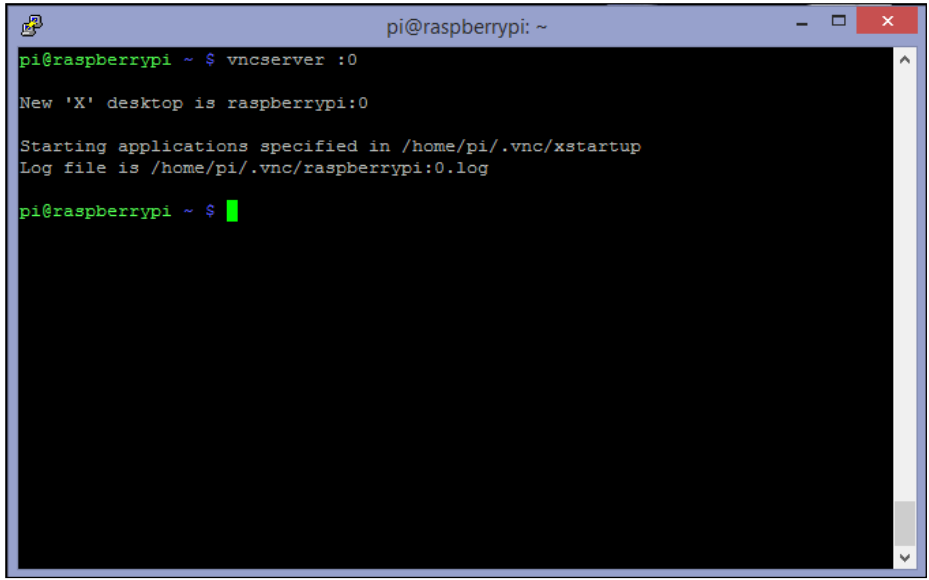

Now we must install in our computer a VNC client, which we will download from the official web site.

http://www.tightvnc.com/download.php

From our computer, we will have to start session of VNCviewer, connecting at the same time to the IP of the Raspberry and the monitor where we have entered the server.

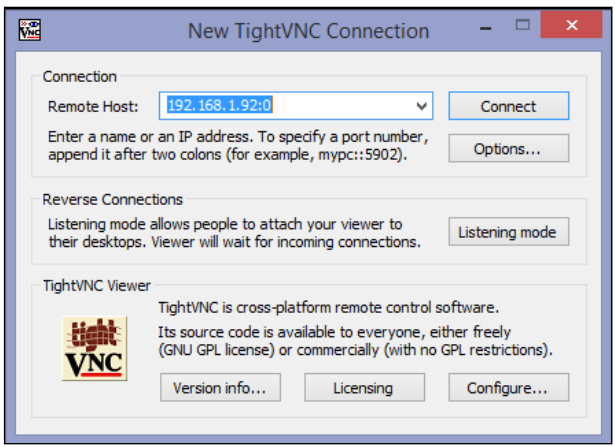

We will be asked for the password previously given to the VNC server.

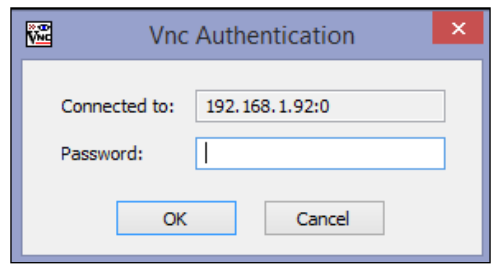

Once we are finished with the authentication process, we will found ourselves on the Raspberry´s desktop.

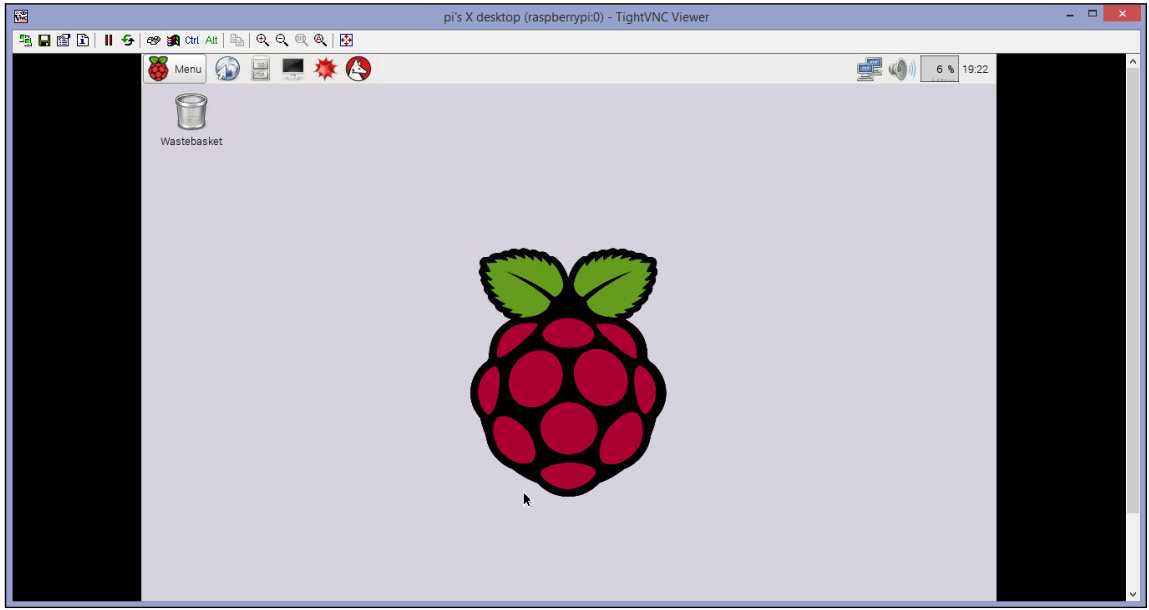

# 3. Defining static IP address for our Raspberry

The DHCP distributes dynamically the addresses to the members of the net, which means that each time a device is connected to that net it will have a different net address[3]. In order to be able to connect to our server (Raspberry) without the need of obtaining the IP address each time, we will give our Raspberry device a static IP address.

To get that we must modify the interface file using the following route:

*/etc/network/interfaces*

First, we must install in our Raspberry a text editor. We will choose the text editor *nano*.

*sudo apt-get install nano*

Creating a backup is recommendable because sometimes the new configuration fails and restoring the default values is needed.

*sudo cp /etc/network/interfaces  /etc/network/interfaces.bk*

Now, the interfaces' file can be modified.

*sudo nano /etc/network/interfaces*

---

[3] DHCP (Dynamic Host Configuration Protocol): this is a protocol of the application level that allows the members of a net to obtain  their setting parameters automatically. A server (usually the entrance door to the net) owns a list of dynamic IPs that are assigned to the clients entering the net.

```
GNU nano 2.2.6                    File: interfaces.bk

auto lo
iface lo inet loopback

auto eth0
allow-hotplug eth0
iface eth0 inet manual

auto wlan0
allow-hotplug wlan0
iface wlan0 inet manual
wpa-conf /etc/wpa_supplicant/wpa_supplicant.conf

auto wlan1
allow-hotplug wlan1
iface wlan1 inet manual
wpa-conf /etc/wpa_supplicant/wpa_supplicant.conf

                        [ Read 17 lines ]
^G Get Help   ^O WriteOut   ^R Read File  ^Y Prev Page  ^K Cut Text   ^C Cur Pos
^X Exit       ^J Justify    ^W Where Is   ^V Next Page  ^U UnCut Text ^T To Spell
```

```
GNU nano 2.2.6                    File: interfaces

#The loopback interface
auto lo
iface lo inet loopback

auto eth0
allow-hotplug eth0
iface eth0 inet static
#Static IP
address 192.168.1.92
#Gateway IP
gateway 192.168.1.1
netmask 255.255.255.0
#Network address
network 192.168.1.0
broadcast 192.168.1.255

auto wlan0
allow-hotplug wlan0
iface wlan0 inet manual
                        [ Read 26 lines ]
^G Get Help   ^O WriteOut   ^R Read File  ^Y Prev Page  ^K Cut Text   ^C Cur Pos
^X Exit       ^J Justify    ^W Where Is   ^V Next Page  ^U UnCut Text ^T To Spell
```

The last image shows us how the Raspberry interface's parameters are configured to set a fixed IP direction in our server.

Rebooting the network is needed to finalize the process. *Networking* file is in charge of enabling/disabling net interfaces, this file with many others are located in the */etc/init.d* directory and they are executed when the Raspberry initializes season. For this reason the network is rebooted.

*sudo /etc/init.d/networking restart*

Files inside the directory  */var/lib/dhcp* are deleted because all of them contains addresses used by the Raspberry Pi.

*sudo rm /var/lib/dhcp/\**

Once it is done, we reboot the Raspberry.

*sudo reboot*

Verifying the interfaces ´configuration.

*ifconfig*

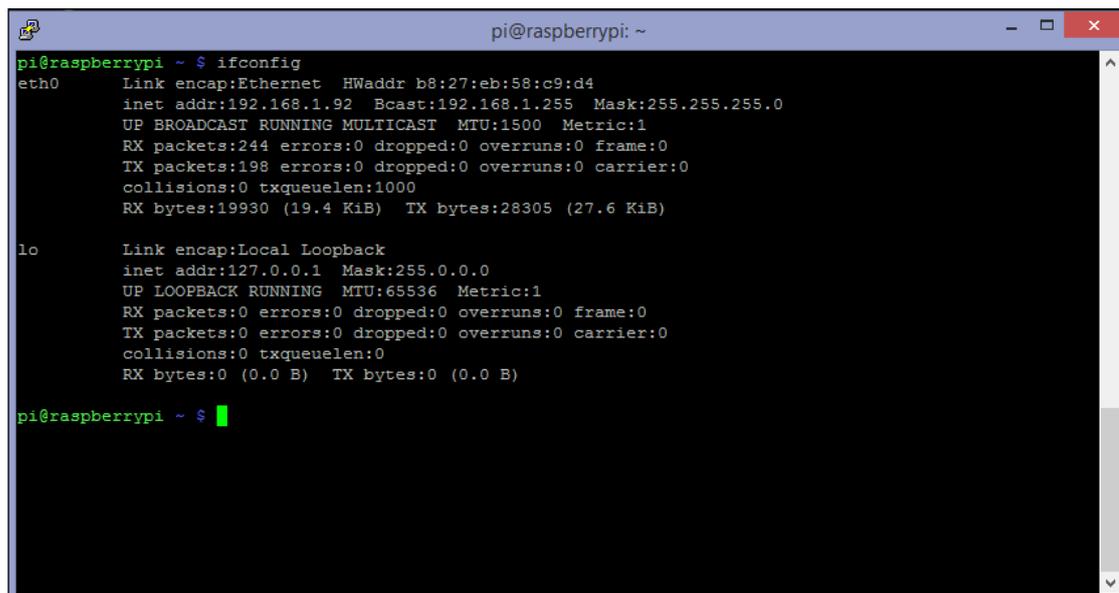

It could be seen that the Raspberry IP address is the expected one, and for confirming that we are still inside the network many pings to the router are executed.

*ping 192.168.1.1 –c10*

```
                                      pi@raspberrypi: ~                    _  □  ✕
pi@raspberrypi ~ $ ping 192.168.1.1 -c10
PING 192.168.1.1 (192.168.1.1) 56(84) bytes of data.
64 bytes from 192.168.1.1: icmp_req=1 ttl=64 time=0.595 ms
64 bytes from 192.168.1.1: icmp_req=2 ttl=64 time=0.519 ms
64 bytes from 192.168.1.1: icmp_req=3 ttl=64 time=0.493 ms
64 bytes from 192.168.1.1: icmp_req=4 ttl=64 time=0.470 ms
64 bytes from 192.168.1.1: icmp_req=5 ttl=64 time=0.460 ms
64 bytes from 192.168.1.1: icmp_req=6 ttl=64 time=0.454 ms
64 bytes from 192.168.1.1: icmp_req=7 ttl=64 time=0.507 ms
64 bytes from 192.168.1.1: icmp_req=8 ttl=64 time=0.475 ms
64 bytes from 192.168.1.1: icmp_req=9 ttl=64 time=0.420 ms
64 bytes from 192.168.1.1: icmp_req=10 ttl=64 time=0.469 ms

--- 192.168.1.1 ping statistics ---
10 packets transmitted, 10 received, 0% packet loss, time 8994ms
rtt min/avg/max/mdev = 0.420/0.486/0.595/0.047 ms
pi@raspberrypi ~ $
```

All the pings have been received, then the server definitely has an static IP.

# 4. Accessing from outside the local network

The purpose of this server is offer support to a mobile phone application; databases and web page. The server has to be accessible from outside our local network, in the next points we explain how to achieve it.

## 4.1. Port Forwarding

All the request connections done to our public IP (external router address) are forward to a concrete port and to the Raspberry in our local network. This technique is known as *Port Forwarding* and the corresponding changes can be executed on the router.

The request demand web traffic, the port assigned by the *IANA*[4] to web traffic is the 80. Additionally, we would like to connect remotely to the server. SSH traffic port is 22, then all the traffic in this port is forwarded to the address of our Raspberry device.

Nombre del Servidor:
- ○ Selecciona un Servicio: Selecciona Uno ▼
- ● Servidor Personalizado: Raspberry - MobilitApp

Dirección IP del Servidor: 192.168.1.92

Salvar/Aplicar

| Puerto Externo Inicial | Puerto Externo Final | Protocolo | Puerto Interno Inicial | Puerto Interno Final |
| --- | --- | --- | --- | --- |
| 22 | 22 | TCP ▼ | 22 | 22 |
| 80 | 80 | TCP ▼ | 80 | 80 |
|  |  | TCP ▼ |  |  |
|  |  | TCP ▼ |  |  |
|  |  | TCP ▼ |  |  |
|  |  | TCP ▼ |  |  |
|  |  | TCP ▼ |  |  |
|  |  | TCP ▼ |  |  |
|  |  | TCP ▼ |  |  |
|  |  | TCP ▼ |  |  |
|  |  | TCP ▼ |  |  |
|  |  | TCP ▼ |  |  |

Testing the mentioned ports for the IP 192.168.1.92 is required.

*telnet [dirección IP] [puerto]*

The test is done for ports 80 and 22.

---

[4] IANA (Internet Assigned Numbers Authority): entity which supervises global IP assignations, domain name servers, traffic in the ports from 0 to 1023, and other resources related with internet protocols.

## 4.2. Dynamic DNS

An ISP[5] server set the external address of our router every time it is turned on. This will cause problems if we want to access to the server from the outside because we would need to know the router IP and it will be changing.

Dynamic DNS refresh our IP when it changes. The solution used is the services provided by no-ip, which allow us to create a host with the desired domain and name. This host will play the role of a DNS.

In the no-ip web site we choose *add host* option, and set a name and a domain and define which kind of host we want, in our case it is a type A[6] host.

In the Raspberry will be installed the client part of the DNS service, this one monitores the external IP and when it changes it will notify to the DNS server. Then the server refreshes the address.

These are the followed steps:

Go to the directory where the software will be installed and download it:

*cd /usr/local/src*

*sudo wget http://www.no-ip.com/client/linux/noip-duc-linux.tar.gz*

When it finishes decomprise the programme, enter in the recently created no-ip folder and start installation:

*sudo tar –xvzf noip-duc-linux.tar.gz*

---

[5] ISP(Internet Service Provider): the enterprise which provides internet connection.
[6] Tipo A: used for assigning domains to a concrete IP v4.

*cd no-ip-2.1.9-1*

*sudo make install*

No-ip name account and password are saved in the route ***/usr/local/etc/no-ip2.conf***.

The ***/usr/local/bin/noip2*** file executes the host refresh, and what we want it is a refresh everytime the server restarts. In the directory ***/usr/local/src/noip-2.1.9-1*** there is some information about how to proceed.

Firstly, we create the noip2 file in the **/etc/init.d** route with the following content:

*sudo nano /etc/init.d/noip2*

```
##################################################################
#!/bin/sh
#./etc/rc.d/init.d/functions #descomente/modifique para su
killproc
case "$1" in
start)
echo "Iniciando noip2..."
/usr/local/bin/noip2
;;
stop)
echo -n "Apagando noip2..."
killproc -TERM /usr/local/bin/noip2
;;
*)
echo "Uso: $0 {start|stop}"
exit 1
esac
exit 0
#################################################################
```

Give permissions to the script to allow its execution and update the starting configuration.

*sudo chmod 755 /etc/init.d/noip2*

*sudo update-rc.d noip2 defaults*

Restart the Raspberry and verify if the process is executed. The command *pstree* shows us the processes in execution.

## 4.3. Security in the Server

The server is now vulnerable because it could be attacked for third parties; therefore we are going to increase our security in the server side.

**Crate a new user and change the default pi password**

New user´s creation

*sudo adduser newuser*

A new password is required; try to find a robust password.

Set the same permissions as pi user.

*sudo visudo*

Administrator permissions are given using *sudoers.d* file.

For changnig the pi user password we should initialize session with it and then write the command *passwd*. Then the old and new passwords are asked for.

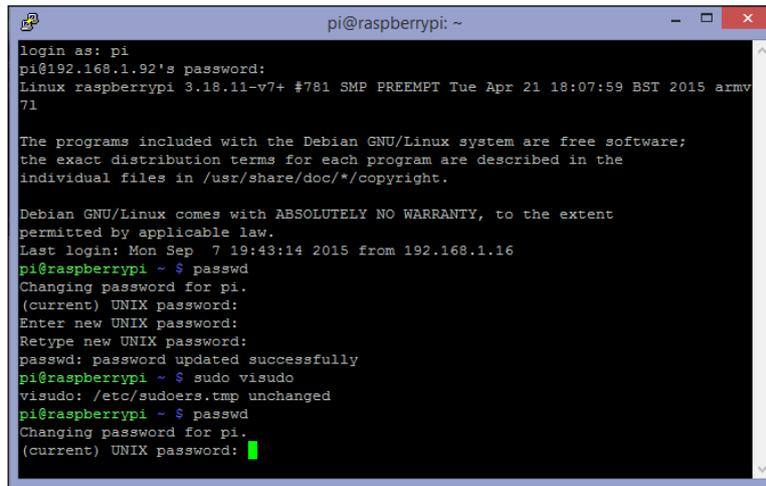

Every time we want to change the password we should follow the lasts steps. It is recommended changing passwords frequently.

**SSH Configuration**

Changing the SSH port is a good security measure.

Open the configuration file to edit its values:

*sudo nano /etc/ssh/sshd_config*

Service SSH ports              [6783-6785]

Another possible change is disabling the SSH connection to the user root.

Not allow the SSH connection to the user root    PermitRootLogin no

# 5. LAMP Configuration

Our server provides web page and database services. LAMP configuration is a common solution for server like the explained in this annex. The Operating System used is Raspbian, from the Linux family.

Apache, MySQL and PHP can be downloaded from the Raspbian repositories.

Update repositories:

*sudo apt-get update*

**Apache**

Apache is an HTTP server, which provides us the web service.

Install Apache:

*sudo apt-get install apache2*

**MySQL**

MySQL is a database server and it will be the keep the application users' database.

Install MySQL:

*sudo apt-get install mysql-server*

**PHP5**

Install PHP and MySQL libraries which allow PHP codes accesing the MySQL database.

Install PHP5:

*sudo apt-get install php5*

*sudo apt-get install php5-mysql*

Finally, we create a file to confirm the activity of the web server and PHP libraries; it is saved in *var/www/* directory with *.php* extension and the following content:

```
<?php
  print <<< EOT
<!doctype html>
<html lang="en">
<head>
<meta charset="UTF-8">
<title>Test successful</title>
</head>
<body>
<h1>Test successful</h1>
<p>Congratulations.</p>
<p>Your webserver and PHP are working.</p>
</body>
</html>
EOT;
?>
```

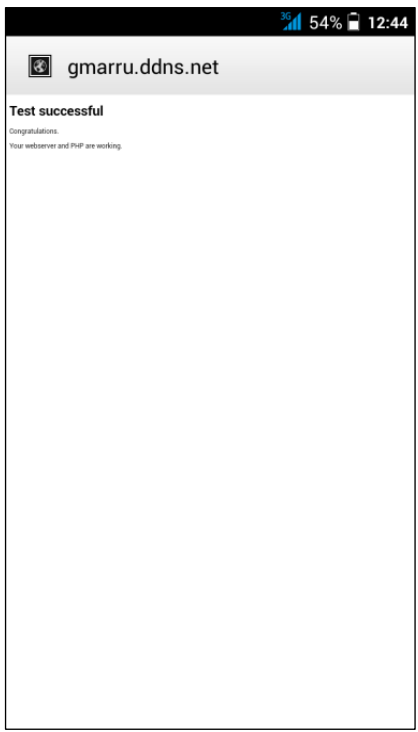

We access to our host from a device outside the local network, for example a mobile phone.

If the result is like the shown in the figure it means that the LAMP configuration progress has been successful.

In addition, with this last step we verify external Access to our server.

# Annex B – Extra Features

Some extra features have been made to complete the project. The objective of these features is to move MobilitApp closer to the citizens.

## Web Page

Having its own web site allows sharing the application through the Internet and present MobilitApp to the world. On the web page a brief project explanation and some specifications about app functionalities could be found.

*http://mobilitapp.noip.me/*

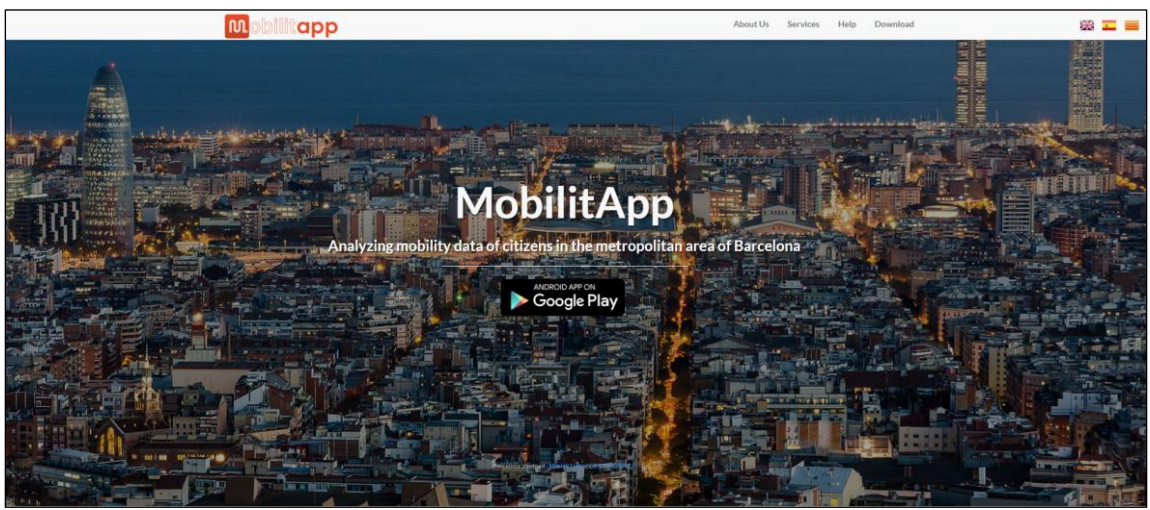

**Mobility Information**

MobilitApp records synchronously and while running in the background, periodic location updates from its users. The information obtained is processed to analyze the mode of transportation used by the user. At the moment we are able to recognise: bus, car, metro, train, bicycle and on foot.

Users are then able to visualize on a map, their locations history day by day. The user's information is treated completely anonymous.

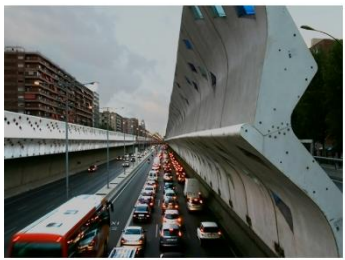
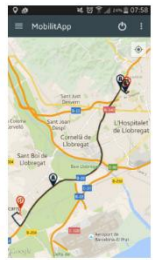
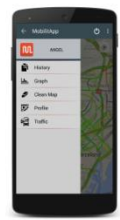
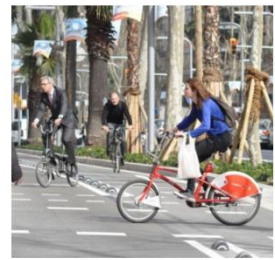

## Real-Time Traffic State and Incidences

MobilitApp integrates real-time traffic state information provided by the Barcelona City Hall and the traffic incidence information provided by the Spanish Traffic Authority: Dirección General de Tráfico.

## Transportation analytics for ATM, the transport authority of the metropolitan area of Barcelona

We provide tools to filter and analize citizens mobility patterns.
In collaboration with the Metropolitan Transport Authority of Barcelona.

## Help

Some hints to use the MobilitApp.

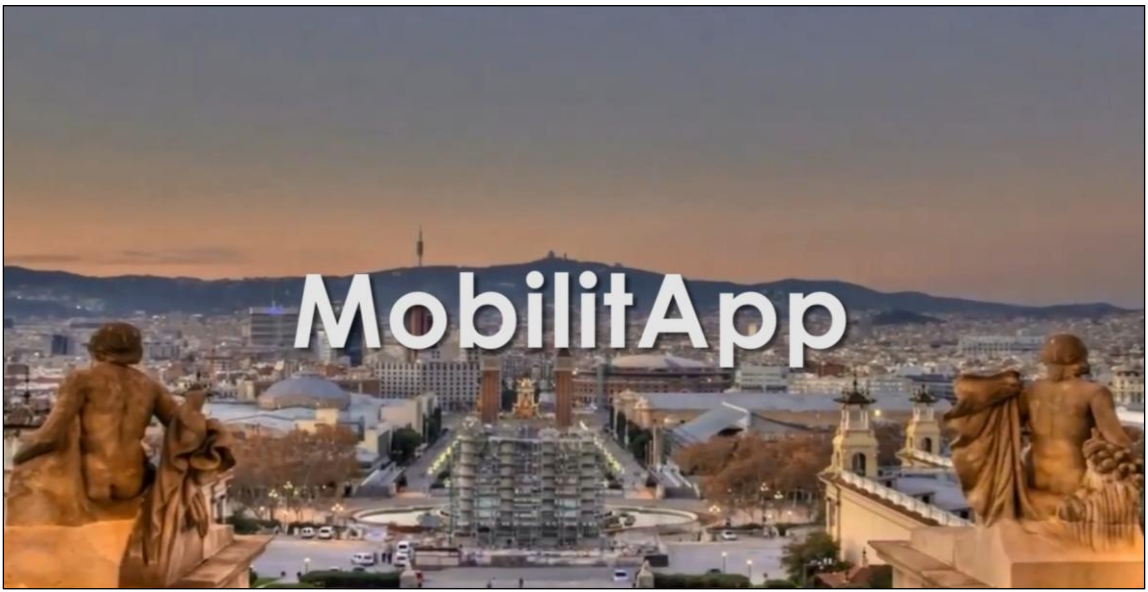

Who we are?

The MobilitApp is being developed by researchers and students of the INRISCO Spanish project, at the Dept. of Telematics Engineering of the Universitat Politècnica de Catalunya under the supervision of prof. Mónica Aguilar and PhD student Silvia Puglisi.

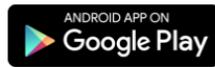

Developers

**Marc Llahona**         **Sergi Casanova**         **Ángel Torres**
         **Gerard Marrugat**         **Antonio Cárdenas**

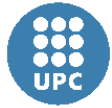 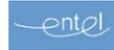

Barcelona 2016

## Promotional Video

A promotional video has been edited to promote MobilitApp among the Barcelona citizens. The video has been film by the development team as well as the composition of the soundtrack. This video can be watched at:

https://www.youtube.com/watch?v=asVWO0HnvOM

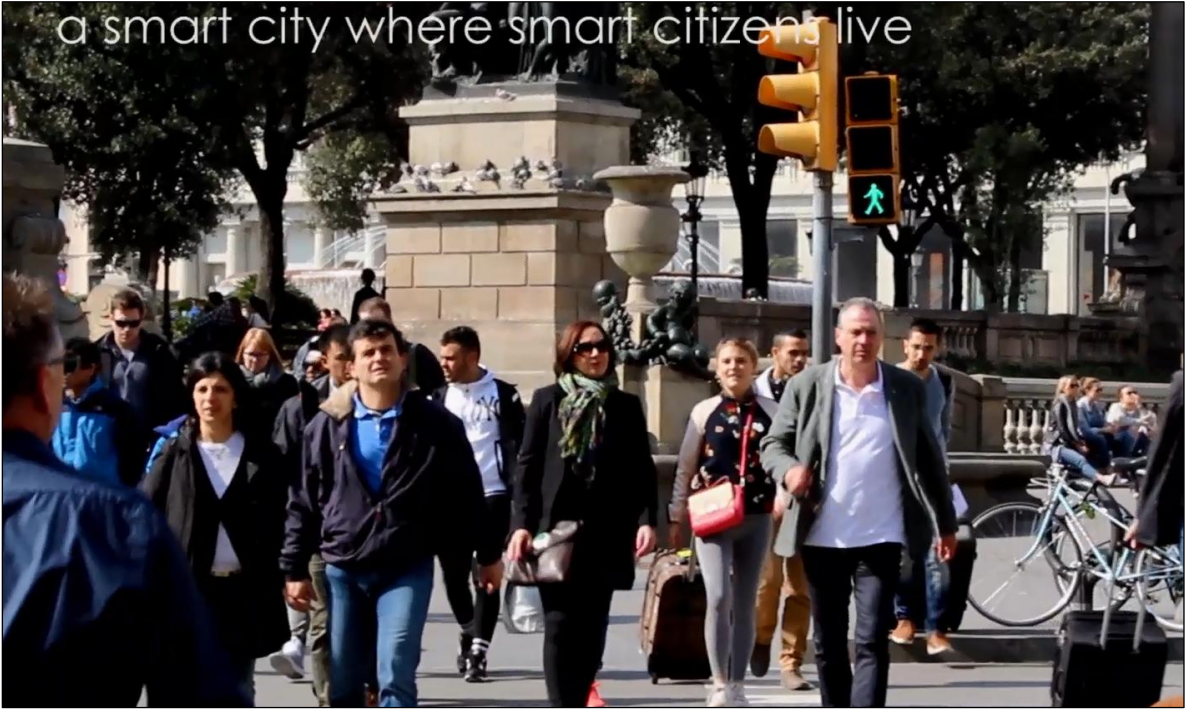

a smart city where smart citizens live

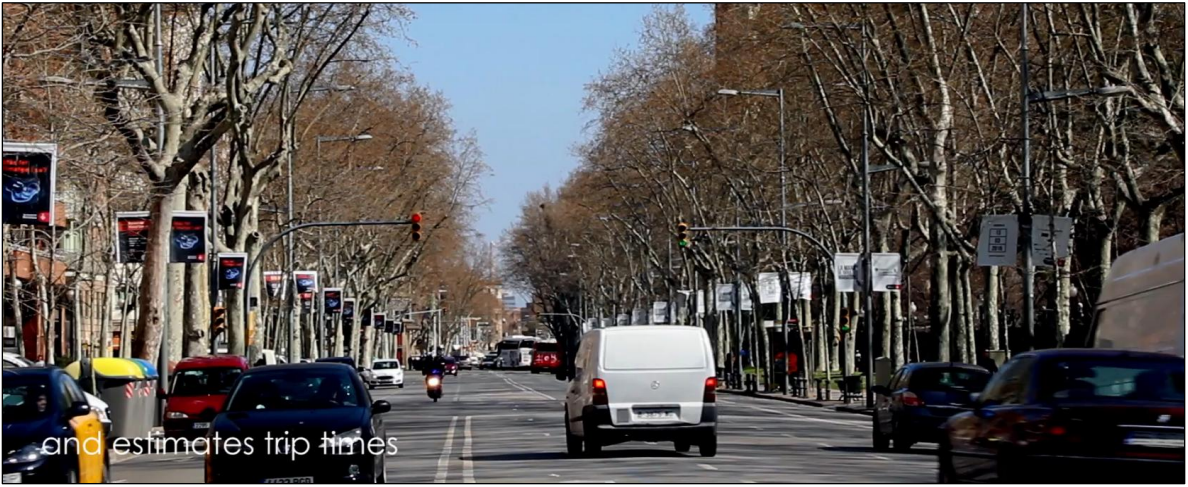

and estimates trip times

# Annex C – Android Phones, Tablets and Watches which do not receive accelerometer events in standby (screen off)

During the Accelerometer Sensor Listener design we have found an important issue for the developers' community. There is a list of Android devices which cannot read from the accelerometer while its screen is off. It is a *bug* (software error) of the hardware manufacturer. Here is the list:

- Alcatel Move
- Archos 101 tablet
- Asus Fonepad 7
- Asus Memo Pad 7 HD
- Fairphone 1
- Google Nexus One
- Google Nexus 6
- HTC Aria
- HTC Desire
- HTC Desire HD
- HTC Desire S
- HTC Desire Z
- HTC Droid Eris
- HTC Droid Incredible
- HTC Evo
- HTC Evo 4G
- HTC G1
- HTC Hero
- HTC Incredible
- HTC Legend
- LG Phoenix
- Mediasonic MTP-710
- Motorola Atrix 4G
- Motorola Atrix MB860
- Motorola Droid
- Motorola Droid X2

- HTC Magic
- HTC One-M8
- HTC One-S
- HTC Rezound
- HTC Tattoo
- HTC Thunderbolt
- Huawei Ascend g300
- Huawei Sonc U8650
- Lenovo a3000 Ideatab
- Lenovo a369i
- Lenovo s820
- Lenovo K900
- LG G100 Smartwatch
- LG Optimus 2X
- LG Optimus Chic E720
- LG Optimus L5 II
- LG Optimus S
- LG Optimus T
- LG P970
- LG P990

- Motorola Milestone
- Motorola Photon 4G MB855
- Motorola Razr Maxx
- Motorola Razri
- Samsung Galaxy Ace
- Samsung Galaxy Ace 2
- Samsung Galaxy Ace 3
- Samsung Galaxy Mini
- Samsung Galaxy Note
- Samsung Galaxy Note 2
- Samsung Galaxy S Duos
- Samsung Galaxy S2
- Samsung Galaxy Xcover 2
- Samsung Moment
- Sharp IS01
- Sharp Lynx
- Sony Xperia X10
- T-Mobile 2X
- T-Mobile G-Slate
- T-Mobile G1
- T-Mobile MyTouch 3G Slide
- ZTE Z667T